\documentclass{aa}  

\usepackage{graphicx}
\usepackage{txfonts}
\usepackage{lipsum}
\usepackage{subcaption}         
\usepackage{lscape}             
\usepackage{placeins}           
                                
\usepackage{txfonts}
\usepackage[colorlinks=true, citecolor=blue, urlcolor=blue]{hyperref}
\usepackage{placeins}
\usepackage{color, soul}
\usepackage{ulem}
\usepackage{natbib}
\usepackage{lipsum}
\bibpunct{(}{)}{;}{a}{}{,} 
\usepackage{tabularx}  
\usepackage{caption}

\begin{document}

   \title{Probing the origin of the LHAASO Peanut with a hidden central accelerator}


%
%
%

    \titlerunning{Probing the LHAASO Peanut}

   \author{Agnibha De Sarkar
          }

   \institute{Institute of Space Sciences (ICE, CSIC), Campus UAB, Carrer de Can Magrans s/n, E-08193 Barcelona, Spain \\
             \email{desarkar@ice.csic.es}\\
            }
    
   \date{Received XXXX; accepted YYYY}
 
\abstract{A highly asymmetric, giant Peanut-shaped structure has been detected by the Large High Altitude Air Shower Observatory (LHAASO) in ultra-high-energy gamma rays at high Galactic latitude, $b\approx-17.5^{\circ}$, suggesting anisotropic particle transport around a Galactic accelerator.
We {investigate} whether its spectrum and morphology can arise from a hidden central accelerator embedded in a projected anisotropic halo.
The baseline model accounts for the diffuse strip and central excess, while a phenomenological prescription for a localized magnetic field reorientation toward the line of sight (LOS), {together with a particle-conserved remapping of the electrons downstream of the far excess into an emitting column of prescribed transverse extent, enhances the emission} at the far excess position.
{We explore various source ages, along with different combinations of ambient magnetic field strength and parallel diffusion coefficient, while accounting for radiative cooling when estimating the viewing geometry.}
{The exploration reveals that relatively rapid parallel transport enhances the far component while preserving a dominant diffuse strip. Representative configurations remain compatible with the observed spectral energy distribution, although the fitted far component remains comparatively weaker than the central component above 100 TeV, even in cases where their component fluxes are comparable at 25-100 TeV.}
{These results illustrate the plausibility of a hidden central accelerator as the origin of the LHAASO Peanut, while identifying the remaining difficulty in reproducing the relative brightness of its compact components within the adopted phenomenological framework. }
}


   \maketitle
   \nolinenumbers

\section{Introduction}\label{sec:intro}

Extended Tera-electronvolt (TeV) halos around middle-aged pulsars, established around Geminga and Monogem \citep{abey17,albert24}, probe the transport of ultra-relativistic pairs near their sources \citep{sudoh19,giacinti20,fang22,liu22}.
They are commonly modeled with near-source diffusion coefficients suppressed relative to the Galactic average \citep{genolini19,desarkar21}, possibly owing to enhanced ambient or self-generated turbulence \citep{evoli18,schroer22}.
If escaped particles sample many magnetic field coherence lengths before cooling, transport becomes effectively isotropic and the halo appears approximately symmetric. For Geminga, HAWC constrains the coherence length to a few parsecs, with a preferred value of $L_{\rm c}\sim1$ pc \citep{lopezcoto18}.
If instead particles cool before sampling multiple field orientations, or sub-Alfv\'enic turbulence preserves an ordered field, transport becomes anisotropic, with $D_\perp\ll D_\parallel$ and $D_\perp/D_\parallel\sim M_A^4$ for Alfv\'enic Mach number $M_A\ll1$ \citep{liu19}.
The apparent morphology then depends on both the diffusion tensor and viewing geometry, i.e., a magnetic field nearly aligned with the line of sight (LOS) can make the source appear compact, whereas a larger inclination produces an elongated gamma-ray structure.

The Large High Altitude Air Shower Observatory (LHAASO) has revealed a growing population of Galactic ultra-high-energy (UHE) gamma-ray sources emitting above $100$ TeV \citep{1lhaaso24}.
Such sources have traditionally been associated with supernova remnants (SNRs) interacting with molecular clouds or with powerful pulsar wind nebulae (PWNe) \citep[see, e.g.,][]{kar22,desarkar22a,desarkar22b,desarkar23,desarkar24,joshi23,wu23,mitchell24,mitchell24b,abe23,delafuente23}, although massive stellar clusters \citep{haubner25,qiu26,abey21,mitchell26} and microquasars \citep{lhaaso25micro,zhang25,zhang26,abaroa26,kaci25} may also accelerate particles to Peta-electronvolt (PeV) energies.
Pulsar halos have likewise emerged as possible contributors to the UHE population \citep[see, e.g.,][]{zheng23, martin25, adams25,desarkar26,fang26,wei26}, but their identification in the crowded Galactic plane is complicated by source confusion and by contamination from nearby PWNe, SNRs, molecular clouds, and Galactic diffuse emission \citep{mestre22}.

A particularly unusual identification made by LHAASO is the giant Peanut-shaped UHE gamma-ray structure at $b\approx-17.5^\circ$, extending over $l\approx135^\circ-140^\circ$ and oriented at $13.8^\circ\pm1.2^\circ$ \citep{lhaaso25}. It comprises a remarkably straight diffuse strip and two aligned compact excesses. The strip measures $4.60^\circ\pm0.44^\circ$ by $0.45^\circ\pm0.12^\circ$, corresponding to $\sim250 \, {\rm pc}\times25 \, {\rm pc}$ at $3.15$ kpc, and contributes $\approx$ $70\%$ of the total flux, while LHAASO J0207+4300 (J0207) near the center and LHAASO J0216+4239 (J0216) near one end remain visible to several hundred TeV. Interestingly, the highest-energy photon of $760^{+60}_{-40}$ TeV, was detected near J0207 \citep{lhaaso25}. The only known energetic object in the region is the millisecond pulsar (MSP) PSR J0218+4232 near J0216, with $P_{\rm MSP}\approx2.3$ ms \citep{navarro95,bassa03}, $\dot P_{\rm MSP}\approx7.74\times10^{-20} \, {\rm s\,s^{-1}}$, $\dot E_{\rm MSP}\approx2.44\times10^{35} \, {\rm erg\,s^{-1}}$ \citep{navarro95,gotthelf17,abdo10,deneva19,acciari21}, $\tau_{\rm MSP}\approx4.8\times10^8$ yr \citep{gotthelf17}, $B_{\rm MSP}\approx4.3\times10^8$ G \citep{konar17}, and a distance of $\sim3.15$ kpc \citep{verbiest14}. Its proper motion, $\mu_{\rm MSP}\approx6.5 \ {\rm mas\,yr^{-1}}$, is approximately parallel to the strip \citep{du14}. The nearby pulsars PSR J0211+4235 and PSR J0209+4331 are too weak to be compelling counterparts \citep{lhaaso25}.

The LHAASO morphology was modeled above $25$ TeV as a rectangular strip plus two point-like hotspots and shows no significant change between the $25-100$ TeV and $>100$ TeV bands \citep{lhaaso25}. The lack of compact molecular clouds and the presence of relatively uniform atomic gas in the region disfavor a hadronic scenario, while favoring a leptonic interpretation, where the inverse-Compton (IC) scattering of Cosmic Microwave Background (CMB) photons by relativistic electrons causes the gamma-ray emission \citep{lhaaso25}. 
Although anisotropic diffusion can, in principle, explain the elongated strip, the hotspots were introduced as independent spatial templates. Moreover, the role of PSR J0218+4232 as the dominant origin remains uncertain, given the short PeV electron cooling time with respect to its age, its insufficient proper motion displacement, and the acceleration, confinement, and escape limitations of a possible associated bow shock pulsar wind nebula (BSPWN). We therefore {employ a simple phenomenological model to test} whether particle injection from an unresolved pulsar-PWN system near the central excess, followed by anisotropic diffusion and a phenomenological LOS-aligned effective particle column at the far hotspot, {resulting from particle-conserved remapping of the downstream electrons}, can reproduce the observed spectral energy distribution (SED) of the Peanut and its diffuse, central, and far components. {We have explored the dependence of the model results on source age, ambient magnetic field strength, and parallel diffusion coefficient, to examine whether the component emission signatures can be reproduced self-consistently in our suggested picture.}
{In Sect. \ref{app: MSP}, we have discussed whether PSR J0218+4232, or its possible associated BSPWN, can be the origin of the Peanut. In Sect. \ref{sec:model}, we have provided the details of the model employed in this paper. Sect. \ref{sec:results} reports the results of the model, and Sect. \ref{sec:conclusion} concludes.}

\section{Can PSR J0218+4232 or its possible BSPWN power the Peanut?}
\label{app: MSP}

PSR J0218+4232 is the only known energetic pulsar projected close to the Peanut and is therefore the most obvious counterpart to test \citep{lhaaso25}. The relevant question, however, is not whether this moving MSP could host a compact BSPWN, but whether the MSP or its BSPWN can supply the PeV particles responsible for the entire structure. We find this possibility rather demanding. This conclusion does not argue against anisotropic transport as the origin of the morphology, but instead favors an accelerator located closer to the center of the Peanut.

In a leptonic interpretation, photons extending to $\sim0.7$-$0.8\,\mathrm{PeV}$ imply parent electrons of order PeV, with a representative energy of approximately $1.7\,\mathrm{PeV}$ for IC scattering on the CMB in the estimate adopted by the LHAASO analysis \citep{lhaaso25}.
For an isotropic pitch angle distribution, their synchrotron cooling time in {a representative} 1 $\mu$G {ambient} magnetic field is
\begin{equation}
t_{\rm syn}
=
\frac{6\pi m_{\rm e}c}
{\sigma_{\rm T}B^2\gamma}
\approx
1.25\times10^4
\left(\frac{E_{\rm e}}{1 \ {\rm PeV}}\right)^{-1}
\left(\frac{B}{1 \ \mu{\rm G}}\right)^{-2}
 {\rm yr}.
\label{eq:tsyn_pev}
\end{equation}
This is many orders of magnitude shorter than the characteristic age of the MSP. The relevant particles, therefore, cannot be a fossil population accumulated over the MSP lifetime and must instead have been injected continuously or within approximately their most recent cooling time.

The source morphology also disfavors a proper motion origin. At a distance $d=3.15$ kpc, the observed size of $0.45^\circ\times4.6^\circ$ corresponds to approximately $25\times250$ pc. The measured proper motion, $\mu\approx6.5 \ {\rm mas \ yr^{-1}}$, implies a transverse velocity of approximately $90-100 \ {\rm km \ s^{-1}}$ \citep{verbiest14}. During the cooling time of a PeV electron, the pulsar moves by only
\begin{align}
\Delta\theta
&\approx
\mu t_{\rm syn}
\nonumber\\
&\approx
2.3\times10^{-2}
\left(\frac{\mu}{6.5 \ {\rm mas \ yr^{-1}}}\right)
\left(\frac{E_{\rm e}}{1 \ {\rm PeV}}\right)^{-1}
\left(\frac{B}{1 \ \mu{\rm G}}\right)^{-2}
 {\rm deg}.
\end{align}
This corresponds to only $\approx1.2$ pc, decreasing to $\approx0.7$ pc for $E_{\rm e}=1.7$ PeV, and is negligible compared with the $\sim250$ pc length of the Peanut. A direct proper motion trail is therefore implausible, consistent with previous traveling-PWN modeling under isotropic and homogeneous diffusion conditions \citep{xie25}.

One could alternatively argue that the relevant accelerator might be a BSPWN created by the motion of PSR J0218+4232 through the ambient medium. A BSPWN could, in principle, provide a compact acceleration site, after which the escaping particles are redistributed by the large-scale magnetic field. Its characteristic stand-off distance follows from the ram pressure balance
\begin{align}
R_0
&=
\left(
\frac{\dot E}
{4\pi c\rho v_{\rm psr}^2}
\right)^{1/2}
\nonumber\\
&\approx
1.7\times10^{-2}
\left(
\frac{\dot E}
{2.4\times10^{35} \ {\rm erg \ s^{-1}}}
\right)^{1/2}
n_0^{-1/2}
v_{100}^{-1}
\ {\rm pc},
\label{eq:stand_off}
\end{align}
where $n_0=n/(1 \ {\rm cm^{-3}})$,
$v_{100}=v_{\rm psr}/(100 \ {\rm km \ s^{-1}})$, and
$\rho\approx1.4m_{\rm p}n$.
At $d = 3.15$ kpc, this corresponds to
\begin{equation}
\theta_0
\approx
1.1^{\prime\prime}
\left(\frac{d}{3.15 \ {\rm kpc}}\right)^{-1}
n_0^{-1/2}
v_{100}^{-1}.
\end{equation}
Even for low halo density, $n=10^{-2} \ {\rm cm^{-3}}$, this corresponds to only $\sim0.17$ pc or $\sim11^{\prime\prime}$. A BSPWN could therefore act only as a compact injector, whereas the large-scale morphology would still have to arise from propagation through the surrounding medium.

{The maximum particle energy in the BSPWN is also constrained by the magnetic confinement scale and the available pulsar wind power.} If a fraction $\eta_B$ of the wind energy flux is magnetic near the bow shock, the magnetic field near the termination region is approximately
\begin{equation}
B_{\rm bs}
\approx
\left(
\frac{2\eta_B\dot E}{cR_0^2}
\right)^{1/2}
\approx
7.6
\left(\frac{\eta_B}{0.01}\right)^{1/2}
n_0^{1/2}
v_{100}
\ \mu{\rm G}.
\label{eq:b_bspwn}
\end{equation}
The corresponding confinement limit is
\begin{equation}
E_{\rm bs}
\sim
eB_{\rm bs}R_0
\approx
0.12
\left(\frac{\eta_B}{0.01}\right)^{1/2}
\left(
\frac{\dot E}
{2.4\times10^{35} \ {\rm erg \ s^{-1}}}
\right)^{1/2}
 {\rm PeV}.
\label{eq:hillas_bspwn}
\end{equation}
This estimate is independent of the ambient density because $B_{\rm bs}\propto n^{1/2}$ and $R_0\propto n^{-1/2}$. For moderate magnetization, the limit is well below the required energy. Even in the extreme limit $\eta_B\rightarrow1$, it reaches only $\sim1.2$ PeV, comparable to or below the electron energy implied by the highest-energy photons.

However, note that for the magnetic field in Eq. \ref{eq:b_bspwn}, the synchrotron loss time inside the BSPWN is
\begin{equation}
t_{\rm syn,bs}
\approx
1.3\times10^2
\left(\frac{E_{\rm e}}{1.7 \, {\rm PeV}}\right)^{-1}
\left(\frac{\eta_B}{0.01}\right)^{-1}
n_0^{-1}
v_{100}^{-2}
\, {\rm yr}.
\end{equation}
This further introduces a potential confinement-escape-cooling tension.
Increasing the magnetization raises the confinement limit only as $\eta_B^{1/2}$, while reducing the synchrotron cooling time as $\eta_B^{-1}$. Particles must therefore remain confined long enough to be accelerated, but escape rapidly enough to avoid synchrotron losses. 

After escape, anisotropic diffusion could still produce the observed elongation. Because PSR J0218+4232 is projected near one end of the structure, particles supplied by it would need to propagate over nearly the full $\sim$250 pc strip length to populate the opposite end. {For an illustrative estimate, we assume that the ordered magnetic field lies approximately in the plane of the sky, so that the projected longitudinal extent directly traces parallel diffusion. Using the synchrotron cooling time for the representative magnetic field of $B=1\,\mu$G, the characteristic diffusion coefficient needed to populate a length $L$ is}
\begin{equation}
D_{\parallel}
\sim
\frac{L^2}{4t_{\rm syn}}
\approx
3.8\times10^{29}
\left(\frac{L}{250 \, {\rm pc}}\right)^2
\left(
\frac{t_{\rm syn}}
{1.25\times10^4 \, {\rm yr}}
\right)^{-1}
\, {\rm cm^2 \, s^{-1}},
\end{equation}
whereas maintaining a transverse half-width $W\approx12.5$ pc requires
\begin{equation}
D_{\perp}
\sim
9.4\times10^{26}
\left(\frac{W}{12.5 \, {\rm pc}}\right)^2
\left(
\frac{t_{\rm syn}}
{1.25\times10^4 \, {\rm yr}}
\right)^{-1}
\, {\rm cm^2 \, s^{-1}}.
\end{equation}
{These estimates give a diffusion coefficient ratio of $D_\parallel/D_\perp\sim400$ in the plane-of-the-sky limit, with both coefficients evaluated at the relevant electron energy. Note that an inclination of the ordered magnetic field toward the LOS would require a larger intrinsic parallel diffusion coefficient to reproduce the same projected length.} {The preceding cooling and propagation estimates use synchrotron losses alone. At a fixed magnetic field, including CMB IC losses with Klein-Nishina (KN) suppression shortens the characteristic cooling time and increases both diffusion estimates. A weaker ambient magnetic field, as considered in our model exploration, reduces synchrotron losses and allows longer propagation. However, CMB IC losses remain operative even in this weak-field regime. Including both processes, the characteristic cooling times remain much shorter than the MSP characteristic age, requiring recent or ongoing injection of the PeV electrons responsible for the present highest-energy emission. Even in the weaker ambient magnetic field, the pulsar displacement over these times remains much smaller than the observed longitudinal extent, making a direct proper motion origin of the structure implausible. The longer cooling time in a weaker ambient magnetic field alone reduces the diffusion coefficients required to populate the observed structure, while leaving the characteristic anisotropy inferred from the $D_\parallel/D_\perp$  ratio unchanged. Anisotropic transport can therefore provide a mechanism for shaping the extended emission. Note that the ambient magnetic field governing propagation is distinct from the local magnetic field near the bow shock. Reducing the former does not, by itself, remove the acceleration, confinement, and escape limitations of the BSPWN scenario. These requirements, arising from the simple estimates discussed above, make the MSP/BSPWN interpretation demanding, and motivate considering an accelerator near the central excess. The detection of the highest-energy photon of $\sim$ 760 TeV near the central excess also supports this motivation.}

\section{The model}
\label{sec:model}

We model the Peanut as IC emission from relativistic electrons continuously injected by an unresolved PWN associated with a misaligned pulsar near the central excess. The injected particles subsequently perform anisotropic diffusion in a large-scale ordered magnetic environment. The elongated strip and central concentration are described with an anisotropic pulsar halo model, while the far excess is {modeled as a projected effective particle column enhancement constructed from the propagated downstream electron population.} The putative pulsar parameters are drawn from the characteristic distribution discussed in \cite{desarkar26a}, and the injected power law spectrum with an exponential cutoff is normalized to the time-dependent spin-down luminosity. The baseline electron distribution is computed with \texttt{PHECT} \citep{fang26phect}, which solves the anisotropic diffusion-loss equation for continuous injection, and is subsequently converted into point spread function (PSF)-convolved IC surface brightness (SB) maps. The diffusion coefficients are parameterized as
\begin{equation}
D_{\perp}(E)=D_{\perp,0}\left(\frac{E}{100 \, {\rm TeV}}\right)^{\delta},
\qquad
D_{\parallel}(E)=M_A^{-4}D_{\perp}(E),
\label{eq:diffusion_coefficients}
\end{equation}
where $M_A$ is the Alfv\'enic Mach number. The ordered magnetic field direction defines the $z$-axis of the transport calculation and is inclined by an angle $\Phi$ to the LOS. 
{To account for radiative cooling when estimating the projected dimensions, we use $t_{\rm eff}=\min[t_{\rm age},t_{\rm cool}(E_*)]$ at a representative electron energy $E_*=400\,\mathrm{TeV}$. Here, $t_{\rm cool}(E)=E/b(E)$, where $b(E)=|dE/dt|$ includes synchrotron and CMB IC losses, with KN suppression included. The approximate projected half-scales are
\begin{equation}
\begin{aligned}
R_{\perp,{\rm obs}}
&\approx 2\sqrt{D_\perp(E_*)t_{\rm eff}},\\
R_{\parallel,{\rm obs}}
&\approx 2\sqrt{
\left[
D_\parallel(E_*)\sin^2\Phi+
D_\perp(E_*)\cos^2\Phi
\right]t_{\rm eff}}.
\end{aligned}
\label{eq:morphology_scaling}
\end{equation}
Thus, the transverse half-width constrains $D_\perp$, whereas the apparent length depends jointly on the diffusion coefficients, source age, radiative cooling, and viewing angle \citep{li26}. These relations provide characteristic geometrical estimates. The electron distribution itself is evolved over the true source age, including continuous injection and energy-dependent losses, whereas the IC emission includes the full electron distribution. Further details of the injection, transport, and radiation calculations are given below.}

\subsection{Particle injection, anisotropic diffusion transport and radiation}\label{app:A}

The hidden pulsar is assumed to inject relativistic electrons continuously over its lifetime. In cylindrical coordinates aligned with the ordered magnetic field, the subsequent transport is governed by the diffusion-loss equation
\begin{equation}
\begin{split}
\frac{\partial N(E,r,z,t)}{\partial t}
&=
\frac{1}{r}
\frac{\partial}{\partial r}
\left[
rD_\perp(E)
\frac{\partial N}{\partial r}
\right]
+
D_\parallel(E)
\frac{\partial^2N}{\partial z^2}
\\
&\quad
+
\frac{\partial}{\partial E}
\left[
b(E)N
\right]
+
Q(E,t)\delta^{(3)}(\boldsymbol{x}),
\end{split}
\label{eq:transport_appendix}
\end{equation}
where $N(E,r,z,t)$ is the differential electron density, $b(E)=|dE/dt|$ is the radiative energy loss rate, and {$Q(E,t)$ is the total differential electron injection rate, and $\delta^{(3)}(\boldsymbol{x})$ represents point source injection at the accelerator position.}
The coefficients $D_\perp$ and $D_\parallel$ are given by Eq. \ref{eq:diffusion_coefficients}. The diffusion tensor is taken to be axisymmetric around the ordered magnetic field.  In cylindrical coordinates aligned with the magnetic field direction, this corresponds to perpendicular diffusion in the radial direction and parallel diffusion along the $z$-axis. For sub-Alfv\'enic anisotropic turbulence, the two coefficients are related through the turbulence in the form of Alfv\'enic Mach number $M_{\rm A}$ ($\ll$ 1), so that the ordered magnetic field direction provides a much faster transport channel than the transverse direction.  The angle $\Phi$ between the $z$-axis and the LOS then fixes how the intrinsic diffusion ellipsoid is projected onto the sky.

The spin-down luminosity of the fiducial pulsar is parameterized as
\begin{equation}
\dot E(t)
=
\dot E_0
\left(
1+\frac{t}{\tau_0}
\right)^{-(n+1)/(n-1)},
\label{eq:spin_down_appendix}
\end{equation}
where $\tau_0$ is the initial spin-down timescale and $n$ is the braking index. For $n=3$, the true age of the pulsar $t_{\rm age}$ is related to $\tau_0$ via
\begin{equation}
\tau_c=t_{\rm age}+\tau_0 .
\end{equation}
where $\tau_c$ is the characteristic age of the pulsar.

The injected spectrum is taken to be a power law with a generalized exponential cutoff,
\begin{equation}
Q(E,t)
=
Q_0(t)
\left(
\frac{E}{E_0}
\right)^{-p}
\exp\left[
-\left(
\frac{E}{E_{\rm c}}
\right)^\beta
\right],
\label{eq:injection_appendix}
\end{equation}
where $p$ is the injection index, $E_{\rm c}$ is the cutoff energy, and the {cutoff-shape parameter} $\beta$ controls the sharpness of the {high-energy suppression.} The normalization is fixed by requiring a fraction $\eta$ of the instantaneous spin-down power to be converted into relativistic pairs
\begin{equation}
\int dE\,E\,Q(E,t)
=
\eta\dot E(t).
\label{eq:injection_normalization}
\end{equation}
For a point source with continuous injection, the solution of Eq. \ref{eq:transport_appendix}, i.e., the electron number density $N(E, r, z)$, illustrates the necessary physical behavior, i.e., particles spread much
further along the ordered field than across it, while radiative cooling restricts the distance that the highest-energy electrons can travel.

For a sky direction \(\boldsymbol{\theta}\), the electron column density is obtained by integration of the electron number density along the LOS,
\begin{equation}
S_e(E,\boldsymbol{\theta})
=
\int N(E,r,z)\,dl .
\end{equation}
The corresponding differential IC surface brightness is
\begin{equation}
s_\gamma(E_\gamma,\boldsymbol{\theta})
=
\frac{1}{4\pi}
\int dE
\int d\epsilon\,
S_e(E,\boldsymbol{\theta})
f(E, \epsilon, E_{\gamma}),
\label{eq:ic_appendix}
\end{equation}
where $f$ denotes the rate at which gamma-ray photons of energy $E_{\gamma}$ are produced through interactions between electrons of energy $E$ and target photons of energy $\epsilon$. The surface brightness profile integrated over an energy band is
\begin{align}
S_{\gamma}(\boldsymbol{\theta})
&=
\int_{E_{\gamma,1}}^{E_{\gamma,2}}
E_{\gamma}\,
s_{\gamma}(E_{\gamma},\boldsymbol{\theta})
\,dE_{\gamma}
\nonumber\\
&=
\int_{\ln E_{\gamma,1}}^{\ln E_{\gamma,2}}
E_{\gamma}^{2}\,
s_{\gamma}(E_{\gamma},\boldsymbol{\theta})
\,d\ln E_{\gamma}.
\label{eq:band_energy_brightness}
\end{align}
This intrinsic profile is convolved with a 2D symmetric Gaussian PSF,
\begin{equation}
\widetilde S_{\gamma, \rm base}
=
G_{\sigma_{\rm PSF}}
*
S_\gamma .
\label{eq:psf_appendix}
\end{equation}
This baseline calculation gives the diffuse anisotropic strip and the central excess associated with the hidden pulsar.

\subsection{Reoriented particle column construction and its emission}
\label{app:reoriented_column}

The formalism discussed in Subsect. \ref{app:A} already calculates the particle density propagated to and beyond the observed far excess position. We further assume that a local change in magnetic field orientation can increase the projected effective particle column at the far excess position using the same downstream population already supplied by the anisotropic halo. {Let $\alpha$ denote the observed angular separation between the central accelerator and the far excess, and let $d$ be the distance to the accelerator. For the field branch directed toward the observer, the distance along the original field to the reorientation point and its distance from the observer are
\begin{equation}
L_{\rm sample}
=
\frac{d\sin\alpha}{\sin(\Phi+\alpha)},
\qquad
d_{\rm bend}
=
\frac{d\sin\Phi}{\sin(\Phi+\alpha)}.
\label{eq:column_observer_geometry}
\end{equation}
The propagated population beginning at $L_{\rm sample}$ and extending downstream supplies the energy-dependent transverse profile of the reoriented component.} We restrict this phenomenological prescription to the favorable configuration in which the locally reoriented field is aligned with the LOS. {Effectively, the full downstream IC-emitting population is remapped into an emitting column with radius $R_{\rm emit}$, preserving its particle number at every electron energy. The column particle normalization is therefore directly determined by the propagated population. Its original contribution is explicitly removed from the baseline emission, thereby avoiding double counting.}

{We denote by $\zeta\equiv|z|$ the distance from the accelerator along the field branch containing the far excess and select the downstream population satisfying $\zeta\geq L_{\rm sample}$, including the boundary. The transverse integration encompasses effectively the full downstream IC-emitting population in the modeled gamma-ray bands. For each electron energy, we define the downstream column density
\begin{equation}
\Sigma_{\rm down}(E,r)=
\int_{\zeta \geq L_{\rm sample}}
N(E,r,\zeta,t_{\rm age})\,d\zeta ,
\end{equation}
where $N$ is the differential electron number density. Thus, $\Sigma_{\rm down}$ is the number of electrons per unit transverse area and per unit electron energy, obtained by integrating along the original
field direction. This column includes the surviving particles throughout the downstream integration domain, with radiative cooling already incorporated in their evolved distribution. The total differential number of electrons in this downstream population is
\begin{equation}
\mathcal{N}_{\rm down}(E)
=
\int_{\mathcal{A}_{\rm down}}
\Sigma_{\rm down}(E,r)\,dA ,
\end{equation}
where $\mathcal{A}_{\rm down}$ denotes the transverse domain included in the downstream integration. For the calculations presented here, this is the circular region encompassing effectively the full
downstream IC-emitting population in the modeled bands. The longitudinal integration extends to the outer boundary of the evolved electron distribution. We place the selected population in an LOS-aligned projected
column with radius $R_{\rm emit}$ centered on the far excess. The column is evaluated at the effective distance $d_{\rm bend}$.

To specify its transverse structure, we uniformly rescale the longitudinally integrated downstream radial profile into the emitting cross-section and normalize it separately at each electron energy. Denoting the resulting profile by $\mathcal{T}(E,r')$, we require
\begin{equation}
\int_0^{R_{\rm emit}}
2\pi r'\mathcal{T}(E,r')\,dr'=1 .
\end{equation}
Here, $r'$ is the distance from the reoriented column axis, and $\mathcal{T}$ has units of inverse area. Its radial and energy dependence are inherited from the integrated downstream distribution. The LOS electron column density is then
\begin{equation}
\Sigma_{\rm emit}(E,r')
=
\mathcal{N}_{\rm down}(E)\mathcal{T}(E,r'),
\end{equation}
with zero column outside $R_{\rm emit}$. Consequently,
\begin{equation}
\int_0^{R_{\rm emit}}
2\pi r'\Sigma_{\rm emit}(E,r')\,dr'
=
\mathcal{N}_{\rm down}(E),
\end{equation}
so the downstream particle number is conserved at every electron energy. Note that the particle-conserved remapping described above is prescribed phenomenologically, and the electron energies are retained from the evolved transport solution. The column normalization is therefore fixed by the downstream population.}

The IC emissivity of the reoriented component is calculated using the CMB with $T_{\rm CMB}=2.73$ K and $u_{\rm CMB}=0.26 \, {\rm eV\,cm^{-3}}$. The target photons are described by the Planck distribution,
\begin{equation}
n_{\rm CMB}(\epsilon)
=
C_{\rm CMB}
\frac{\epsilon^{2}}
{\exp\!\left(\epsilon/k_{\rm B}T_{\rm CMB}\right)-1},
\label{eq:planck_cmb}
\end{equation}
where the normalization $C_{\rm CMB}$ is fixed by
\begin{equation}
\int d\epsilon\,
\epsilon\,n_{\rm CMB}(\epsilon)
=
u_{\rm CMB}.
\label{eq:planck_normalization}
\end{equation}

For an electron Lorentz factor $\gamma=E/(m_{\rm e}c^{2})$, we define
\begin{equation}
\Gamma
=
\frac{4\gamma\epsilon}{m_{\rm e}c^{2}},
\qquad
q
=
\frac{E_{\gamma}}
{\Gamma(E-E_{\gamma})}.
\label{eq:kn_variables}
\end{equation}
The isotropic KN differential cross section is
\begin{equation}
\frac{d\sigma_{\rm IC}}{dE_{\gamma}}
=
\frac{3\sigma_{\rm T}}
{4\gamma^{2}\epsilon}
\mathcal{F}(q,\Gamma),
\label{eq:kn_cross_section}
\end{equation}
with
\begin{equation}
\mathcal{F}(q,\Gamma)
=
2q\ln q
+
(1+2q)(1-q)
+
\frac{(\Gamma q)^{2}(1-q)}
{2(1+\Gamma q)} ,
\label{eq:kn_kernel}
\end{equation}
and the kinematically allowed domain
\begin{equation}
\frac{1}{4\gamma^{2}}
\leq q\leq1,
\qquad
E_{\gamma}<E .
\end{equation}
{For a sky direction $\boldsymbol{\theta}$, the projected transverse distance from the modeled far hotspot position is
\begin{equation}
R_\perp(\boldsymbol{\theta})
\sim
d_{\rm bend}
\left|\boldsymbol{\theta}-\boldsymbol{\theta}_{f}\right|,
\end{equation}
where $\boldsymbol{\theta}_{f}$ denotes the angular position of the modeled far hotspot. This relation uses the small-angle approximation within the emitting cross-section.}
The differential IC surface brightness calculated from the emitting electron column is
\begin{equation}
\begin{split}
s_{\gamma,\rm column}
(E_\gamma,\boldsymbol{\theta})
={}&
\frac{c}{4\pi}
\int dE\,\Sigma_{\rm emit}(E,R_\perp)\\
&\times
\int d\epsilon\,
n_{\rm CMB}(\epsilon)
\frac{d\sigma_{\rm IC}}{dE_\gamma}.
\end{split}
\end{equation}
The corresponding energy-band-integrated brightness is
\begin{align}
S_{\gamma,{\rm column}}(\boldsymbol{\theta})
&=
\int_{E_{\gamma,1}}^{E_{\gamma,2}}
E_{\gamma}\,
s_{\gamma,{\rm column}}
(E_{\gamma},\boldsymbol{\theta})
\,dE_{\gamma}
\nonumber\\
&=
\int_{\ln E_{\gamma,1}}^{\ln E_{\gamma,2}}
E_{\gamma}^{2}\,
s_{\gamma,{\rm column}}
(E_{\gamma},\boldsymbol{\theta})
\,d\ln E_{\gamma}.
\label{eq:column_band_brightness}
\end{align}
and the observed column map is obtained after convolution with the Gaussian PSF,
\begin{equation}
\widetilde S_{\gamma,{\rm column}}
=
G_{\sigma_{\rm PSF}}
*
S_{\gamma,{\rm column}}.
\label{eq:column_psf}
\end{equation}

To avoid double counting, we estimate the fraction of the baseline emission produced by the downstream population in the original geometry.
Using the same propagated particle distribution and radiation treatment, we calculate the PSF-convolved surface brightness of the full halo, $\widetilde S_{\gamma,{\rm halo}}^{(0)}$, and of the downstream population, $\widetilde S_{\gamma,{\rm down}}^{(0)}$, where the superscript $(0)$ denotes the original geometry. Their ratio defines the local downstream fraction,
\begin{equation}
f_{\rm down}(\boldsymbol{\theta})
=
\frac{
\widetilde S_{\gamma,{\rm down}}^{(0)}(\boldsymbol{\theta})
}{
\widetilde S_{\gamma,{\rm halo}}^{(0)}(\boldsymbol{\theta})
},
\label{eq:downstream_fraction_appendix}
\end{equation}
where $\boldsymbol{\theta}$ denotes angular position on the sky and $0\leq f_{\rm down}\leq1$. Where the halo surface brightness is negligible, we set
$f_{\rm down}=0$. The corresponding contribution already contained in the baseline is
\begin{equation}
\widetilde S_{\gamma,{\rm rem}}(\boldsymbol{\theta})
=
f_{\rm down}(\boldsymbol{\theta})
\widetilde S_{\gamma,{\rm base}}(\boldsymbol{\theta}).
\label{eq:removed_map_appendix}
\end{equation}

The final PSF-convolved SB map is therefore
\begin{equation}
\widetilde S_{\gamma,{\rm tot}}
=
\widetilde S_{\gamma,{\rm base}}
-
\widetilde S_{\gamma,{\rm rem}}
+
\widetilde S_{\gamma,{\rm column}}.
\label{eq:replacement_map_appendix}
\end{equation}
Thus, the original and reoriented contributions of the downstream population are not retained simultaneously. Particle conservation applies to the complete population included in the construction. {For the source-integrated spectrum, we apply the same component replacement within the source-centered aperture. The baseline, removed, and column spectra are normalized so that their band-integrated fluxes reproduce those of the corresponding PSF-convolved maps within this aperture, independently of the spatial template decomposition we discuss next.}

\subsection{Component decomposition}
\label{app:C}

To quantify the relative contributions of the diffuse strip and the two compact excesses, we perform a simultaneous spatial template decomposition of the model SB maps following the component morphology adopted in the LHAASO MD4 analysis \citep{lhaaso25}. The decomposition consists of a uniform rectangular template describing the elongated strip and two intrinsically point-like templates describing the central and far components. All three templates are convolved with the corresponding instrumental PSF and normalized to unit integral over the sky. The transport calculation is defined in a model frame in which the projected
ordered magnetic field direction is aligned horizontally. We therefore retain the band-dependent MD4 rectangle dimensions, $4.56^{\circ}\times0.68^{\circ}$ and $4.56^{\circ}\times0.40^{\circ}$ in the $25$-$100$ TeV and $>100$ TeV bands, respectively, but center the template on the model origin and align it with the modeled strip. The central and far point source templates are placed at the hidden accelerator and reoriented column positions, respectively, separated by $1.82^{\circ}$. Both are intrinsically unresolved, so their convolved profiles are set entirely by the PSF.

Denoting the PSF-convolved, unit-integral spatial templates of the diffuse strip, central excess, and far excess by $T_{\rm strip}$, $T_{\rm c}$, and $T_{\rm f}$, respectively, we write
\begin{equation}
\widetilde S_{\gamma,{\rm fit}}(\boldsymbol{\theta})
=
F_{\rm strip}T_{\rm strip}(\boldsymbol{\theta})
+
F_{\rm c}T_{\rm c}(\boldsymbol{\theta})
+
F_{\rm f}T_{\rm f}(\boldsymbol{\theta}),
\label{eq:component_decomposition}
\end{equation}
where $\boldsymbol{\theta}$ denotes angular position on the sky and each template satisfies $\int T_i(\boldsymbol{\theta})\,d\Omega=1$. Consequently, $F_{\rm strip}$, $F_{\rm c}$, and $F_{\rm f}$ directly represent the band-integrated fluxes of the three morphological components. {These fluxes are determined simultaneously by minimizing the squared difference between $\widetilde S_{\gamma,{\rm fit}}$ and $\widetilde S_{\gamma,{\rm tot}}$ over the adopted fitting region, subject to $F_{\rm strip},F_c,F_f\geq0$, thereby accounting for the spatial overlap between the diffuse and compact templates. A zero fitted compact component flux means that the adopted decomposition does not require an additional positive point source contribution at that position, and it does not imply an absence of physical emission there.} We also tested the LHAASO MD4 positions and rectangle orientation as a check, but do not use this decomposition quantitatively since our analysis is performed on model SB maps rather than on the LHAASO event, exposure, and background data and therefore does not reproduce the observational likelihood analysis.

\begin{figure*}
    \centering
    \includegraphics[width=0.33\textwidth]{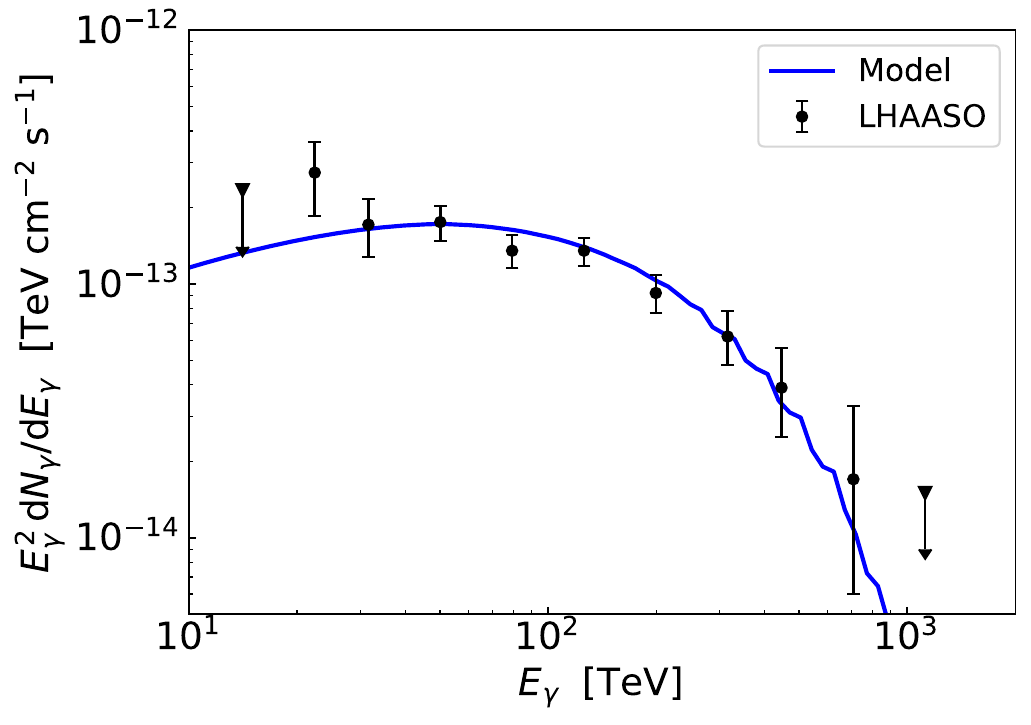}
     \includegraphics[width=0.33\textwidth]{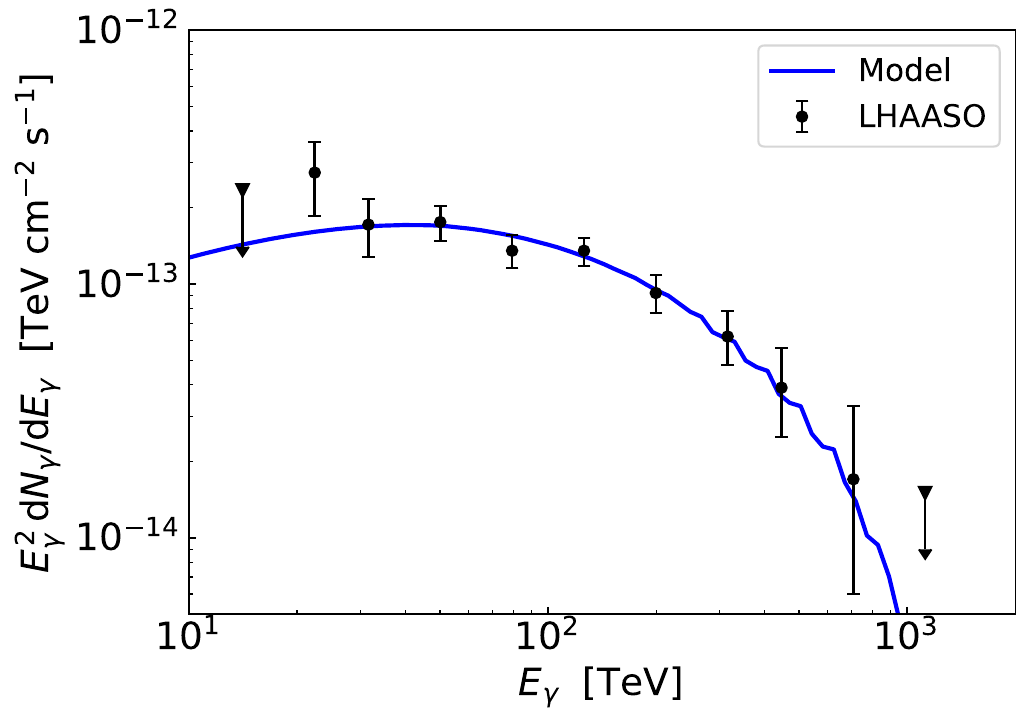}
     \includegraphics[width=0.33\textwidth]{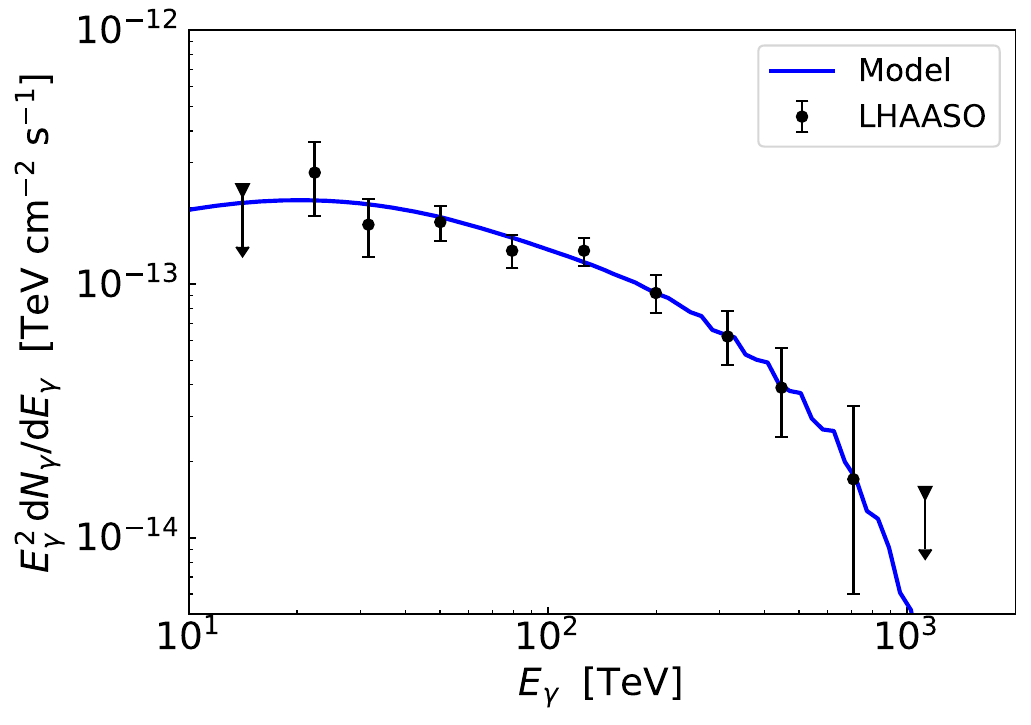}
    \caption{{Spectral energy distribution of the Peanut for Cases 4, 8,
and 12 of Table \ref{tab:parameter_scan}, with source ages
of $10$, $20$, and $30\,\mathrm{kyr}$ from left to right.
The blue curves show the total model emission within the
source aperture, including replacement of the original
downstream contribution by the reoriented column.
Black points show the LHAASO measurements, with downward
arrows indicating upper limits.}}
    \label{fig: spectra}
\end{figure*}

\section{Results}
\label{sec:results}

We begin with the properties of the fiducial hidden pulsar assumed to be located at the center of the Peanut. Motivated by the initial period $P_0$ and the magnetic field $B_0$ distributions discussed in \cite{desarkar26a}, we adopt $P_0 = 0.05$ s and $B_0 = 1.75 \times 10^{12}$ G for the hidden pulsar. One can therefore calculate the initial period derivative to be $\dot{P_0} \approx 6.0 \times 10^{-14} \ \rm s \ s^{-1}$ via the relation $B_0 \approx 3.2 \times 10^{19} \sqrt{P_0 \dot{P_0}}$, assuming the braking index $n = 3$. We can therefore calculate the initial spin-down timescale to be $\tau_0 = P_0/(n-1)\dot{P_0} \approx 1.3 \times 10^4$ yr. {We examine true source ages of $t_{\rm age} =$ $10$, $20$, and $30\,\mathrm{kyr}$ and use the $t_{\rm age} = 20\,\mathrm{kyr}$ case as a representative example.} At this age, the current period and period derivative can be calculated to be $P \approx 0.08$ s and $\dot{P} \approx 3.8 \times 10^{-14} \ \rm s \ s^{-1}$, giving a current spin-down luminosity of $\dot{E} \approx 3 \times 10^{36} \ \rm erg \ s^{-1}$, assuming a moment of inertia value of $I = 10^{45} \ \rm g \ cm^2 $. The characteristic age of the system can thereby be computed as $\tau_c = t_{\rm age} + \tau_0 = P/2\dot{P} \approx 3.3 \times 10^4$ yr. {The hidden pulsar} was assumed to be located at a distance of 3.15 kpc, the same as \cite{lhaaso25}, to remain consistent with the comparison. The consequences of the uncertainty in this distance are discussed in Appendix \ref{app:distance}.

Next, we calculate the required transverse diffusion coefficient $D_{\perp}$ and the angle between LOS and the ordered Galactic magnetic field $\Phi$ by solving the coupled equation given in Eq. \ref{eq:morphology_scaling}. {For each of the three source ages, we explore all combinations of $B=0.1$ and $1\,\mu\mathrm{G}$ with $D_{\parallel,0}=3.2\times10^{29}$ and $3.2\times10^{30}\,\mathrm{cm^2\,s^{-1}}$, yielding twelve configurations. These choices include a magnetic field an order weaker and a parallel diffusion coefficient an order larger than the adopted Galactic average values of $B = 1\,\mu\mathrm{G}$ and $D_{\parallel,0}=3.2\times10^{29} \,\mathrm{cm^2\,s^{-1}}$, respectively, exploring possible conditions in the Galactic halo beyond the Galactic plane. Both diffusion coefficients are normalized at $100\,\mathrm{TeV}$, with a Kolmogorov-like energy dependence, $\delta=0.33$. The perpendicular diffusion coefficient, Alfv\'enic Mach number, and inclination angle are determined jointly from the observed dimensions and the effective propagation time, rather than varied independently. The explored configurations are listed in Table \ref{tab:parameter_scan}.}

{For $400\,\mathrm{TeV}$ electrons, the cooling time is approximately $15\,\mathrm{kyr}$ for $B=1\,\mu\mathrm{G}$ and $28.3\,\mathrm{kyr}$ for $B=0.1\,\mu\mathrm{G}$ under the adopted radiation field. Consequently, the geometrical estimate becomes cooling-limited for the older configurations, depending on the field strength. For the representative configuration, we select $B=0.1\,\mu\mathrm{G}$ and $D_{\parallel,0}=3.2\times10^{30}\,\mathrm{cm^2\,s^{-1}}$, for which $t_{\rm eff}=20\,\mathrm{kyr}$ (see Case 8 from Table \ref{tab:parameter_scan}).} By solving Eq. \ref{eq:morphology_scaling}, we find that to reproduce the {projected half-length} of $\sim 125$ pc (since the accelerator is assumed to be at the center), and the half-width of $\sim 12.5$ pc, with the Peanut being located at a distance of 3.15 kpc and having a true age of $t_{\rm age} =20$ kyr, one would need a perpendicular diffusion coefficient normalization of $D_{\perp, 0} \approx 3.65\times10^{26}\,
\mathrm{cm^2\,s^{-1}}$ at $100\,\mathrm{TeV}$, corresponding to $M_{\rm A} \approx 0.1034$ from the relation given in Eq. \ref{eq:diffusion_coefficients}, and an ordered magnetic field inclination of $\Phi \sim 6.24^\circ$. The large intrinsic parallel diffusion scale is therefore projected onto the sky through the small inclination, while strongly suppressed transverse diffusion maintains the narrow strip.

{The far enhancement is centered at an angular separation $\alpha=1.82^\circ$ from the hidden accelerator. For $d=3.15\,\mathrm{kpc}$ and $\Phi=6.24^\circ$, Eq. \ref{eq:column_observer_geometry} gives $L_{\rm sample}\approx714\,\mathrm{pc}$ and $d_{\rm bend}\approx2.44\,\mathrm{kpc}$. The selected downstream population is redistributed using the particle-conserved remapping prescription discussed in Subsect. \ref{app:reoriented_column}. We adopt $R_{\rm emit}=12.5\,\mathrm{pc}$ for the configurations in Table \ref{tab:parameter_scan} and examine the dependence on this radius separately in Table \ref{tab:emit_radius_scan}. The parameter $R_{\rm emit}$ specifies the outer radius of the remapped transverse profile, rather than its observed PSF-convolved width.} The total model flux and SED are evaluated within a source aperture {of projected radius $155\,\mathrm{pc}$,} enclosing the modeled strip and both compact excesses.

\begin{table*}
\caption{{Explored model parameters and band-integrated energy fluxes of the diffuse strip and the central and far compact components, grouped by source age.}}
\label{tab:parameter_scan}
\centering
\small
\setlength{\tabcolsep}{3pt}
\renewcommand{\arraystretch}{1.2}

\resizebox{\textwidth}{!}{%
\begin{tabular}{ccccccccc|ccc|ccc}
\hline\hline
& \multicolumn{8}{c}{Model parameters}
& \multicolumn{3}{c}{$25$-$100\,\mathrm{TeV}$}
& \multicolumn{3}{c}{$>100\,\mathrm{TeV}$} \\
\cline{1-9}\cline{10-12}\cline{13-15}
Case & $B$ & $D_{\parallel,0}$ & $D_{\perp,0}$ & $\Phi$ & $M_A$
& $t_{\rm eff}$ & $\eta$ & $E_c$
& $F_{\rm strip}$ (\%) & $F_c$ (\%) & $F_f$ (\%)
& $F_{\rm strip}$ (\%) & $F_c$ (\%) & $F_f$ (\%) \\
\hline

\multicolumn{15}{c}{
$t_{\rm age}=10000\,\mathrm{yr}$,\quad
$\tau_c\approx23\,\mathrm{kyr}$,\quad
$\dot E\approx6.3\times10^{36}\,\mathrm{erg\,s^{-1}}$
} \\[2pt]
\hline

1 & 1.0 & 3.2 & 7.305 & 29.118 & 0.2186 & 10.000 & 6.5 & 420
& 20.84 (83.26) & 4.110 (16.42) & 0.07973 (0.32)
& 13.54 (84.24) & 2.317 (14.41) & 0.2169 (1.35) \\

2 & 1.0 & 32.0 & 7.305 & 8.843 & 0.1229 & 10.000 & 6.5 & 420
& 19.61 (78.51) & 3.845 (15.39) & 1.524 (6.10)
& 12.83 (80.07) & 2.126 (13.27) & 1.067 (6.66) \\

3 & 0.1 & 3.2 & 7.305 & 29.118 & 0.2186 & 10.000 & 6.5 & 300
& 20.84 (83.31) & 4.127 (16.50) & 0.04763 (0.19)
& 13.15 (84.43) & 1.990 (12.78) & 0.4339 (2.79) \\

4 & 0.1 & 32.0 & 7.305 & 8.843 & 0.1229 & 10.000 & 6.5 & 300
& 19.61 (78.57) & 3.860 (15.47) & 1.487 (5.96)
& 12.51 (79.80) & 1.849 (11.79) & 1.319 (8.41) \\

\hline
\multicolumn{15}{c}{
$t_{\rm age}=20000\,\mathrm{yr}$,\quad
$\tau_c\approx33\,\mathrm{kyr}$,\quad
$\dot E\approx3.0\times10^{36}\,\mathrm{erg\,s^{-1}}$
} \\[2pt]
\hline

5 & 1.0 & 3.2 & 4.886 & 23.441 & 0.1977 & 14.951 & 7.5 & 800
& 20.03 (82.32) & 3.293 (13.53) & 1.008 (4.14)
& 12.41 (83.36) & 2.307 (15.50) & 0.1706 (1.15) \\

6 & 1.0 & 32.0 & 4.886 & 7.222 & 0.1112 & 14.951 & 7.5 & 800
& 18.90 (75.75) & 3.186 (12.77) & 2.864 (11.48)
& 11.78 (78.61) & 2.129 (14.21) & 1.075 (7.18) \\

7 & 0.1 & 3.2 & 3.652 & 20.114 & 0.1838 & 20.000 & 6.5 & 420
& 20.67 (82.34) & 3.999 (15.93) & 0.4357 (1.74)
& 12.88 (83.11) & 2.223 (14.34) & 0.3947 (2.55) \\

8 & 0.1 & 32.0 & 3.652 & 6.240 & 0.1034 & 20.000 & 6.2 & 420
& 18.41 (75.10) & 3.700 (15.09) & 2.406 (9.81)
& 11.66 (77.12) & 1.987 (13.14) & 1.473 (9.74) \\

\hline
\multicolumn{15}{c}{
$t_{\rm age}=30000\,\mathrm{yr}$,\quad
$\tau_c\approx43\,\mathrm{kyr}$,\quad
$\dot E\approx1.8\times10^{36}\,\mathrm{erg\,s^{-1}}$
} \\[2pt]
\hline

9 & 1.0 & 3.2 & 4.886 & 23.441 & 0.1977 & 14.951 & 13.0 & 850
& 19.41 (81.54) & 3.184 (13.38) & 1.209 (5.08)
& 12.01 (83.18) & 2.339 (16.20) & 0.08975 (0.62) \\

10 & 1.0 & 32.0 & 4.886 & 7.222 & 0.1112 & 14.951 & 13.0 & 850
& 18.35 (75.19) & 3.058 (12.53) & 2.995 (12.27)
& 11.39 (78.67) & 2.147 (14.83) & 0.9408 (6.50) \\

11 & 0.1 & 3.2 & 2.577 & 16.787 & 0.1685 & 28.345 & 10.0 & 480
& 21.29 (81.10) & 4.644 (17.69) & 0.3182 (1.21)
& 12.34 (81.95) & 2.717 (18.05) & 0.0000 (0.00) \\

12 & 0.1 & 32.0 & 2.577 & 5.238 & 0.0947 & 28.345 & 10.0 & 480
& 19.74 (73.02) & 4.496 (16.63) & 2.799 (10.35)
& 11.72 (76.26) & 2.512 (16.35) & 1.135 (7.39) \\

\hline
\end{tabular}%
}

\tablefoot{
Column units and scale factors are: $B$ in $\mu\mathrm{G}$, $D_{\parallel,0}$ in $10^{29}\,\mathrm{cm^2\,s^{-1}}$, $D_{\perp,0}$ in $10^{26}\,\mathrm{cm^2\,s^{-1}}$, $\Phi$ in degrees, $t_{\rm eff}$ in kyr, $\eta$ in units of $10^{-4}$, and $E_c$ in TeV. The Alfv\'enic Mach number $M_A$ is dimensionless. All component fluxes are in units of $10^{-14}\,\mathrm{TeV\,cm^{-2}\,s^{-1}}$. Both diffusion coefficients are normalized at $100\,\mathrm{TeV}$, with $D(E)\propto E^{0.33}$. The effective propagation time is evaluated at an electron energy of $400\,\mathrm{TeV}$ as described in Sect. \ref{sec:model}. All cases use $d=3.15\,\mathrm{kpc}$, $\tau_0\approx13\,\mathrm{kyr}$, $p=1.5$, $\beta=1$, and $R_{\rm emit}=12.5\,\mathrm{pc}$. Component fluxes follow the simultaneous spatial decomposition described in Subsect. \ref{app:C}. Values in parentheses give each component's percentage contribution to the total fitted flux. {For observational comparison, integrating the common shape
cutoff power law component spectra reported in Table 1 of \cite{lhaaso25}, with photon index $\Gamma=1.95$ and cutoff energy $250\,\mathrm{TeV}$, gives $(F_{\rm strip},F_c,F_f) =(16.28\pm1.57,\,3.36\pm0.67,\,2.94\pm0.64)$ in $25$-$100\,\mathrm{TeV}$ and $(10.90\pm1.05,\,2.25\pm0.45,\,1.97\pm0.43)$ in $> 100$ $\mathrm{TeV}$, in the flux units used here. The central and far components correspond to J0207 and J0216, respectively. The corresponding fractions are $(72.12,14.87,13.01)\%$ in both bands because the observational
fit assumes a common spectral shape. The quoted statistical uncertainties propagate the published component normalization errors with that shape held fixed; uncertainty in the common spectral shape and the approximately $10\%$ flux systematic uncertainty are not included.}}
\end{table*}

\begin{table*}
\caption{{Dependence of the fitted diffuse strip, central, and far component fluxes on the emitting column radius for Case 8 of Table \ref{tab:parameter_scan}.}}
\label{tab:emit_radius_scan}
\centering
\small
\setlength{\tabcolsep}{6pt}
\renewcommand{\arraystretch}{1.2}

\begin{tabular*}{\textwidth}{@{\extracolsep{\fill}}c|ccc|ccc@{}}
\hline\hline
& \multicolumn{3}{c}{$25$-$100\,\mathrm{TeV}$}
& \multicolumn{3}{c}{$>100\,\mathrm{TeV}$} \\
\cline{1-4}\cline{4-7}
$R_{\rm emit}$
& $F_{\rm strip}$ (\%) & $F_c$ (\%) & $F_f$ (\%)
& $F_{\rm strip}$ (\%) & $F_c$ (\%) & $F_f$ (\%) \\
\hline

6.25
& 18.38 (74.90) & 3.706 (15.10) & 2.452 (10.00)
& 11.61 (76.69) & 1.993 (13.16) & 1.537 (10.15) \\

12.5
& 18.41 (75.10) & 3.700 (15.09) & 2.406 (9.81)
& 11.66 (77.12) & 1.987 (13.14) & 1.473 (9.74) \\

25
& 18.55 (75.83) & 3.680 (15.04) & 2.233 (9.13)
& 11.82 (78.60) & 1.970 (13.09) & 1.249 (8.30) \\

50
& 18.95 (78.15) & 3.620 (14.93) & 1.677 (6.92)
& 12.16 (82.34) & 1.932 (13.08) & 0.6760 (4.58) \\

\hline
\end{tabular*}

\tablefoot{
{The emitting column radius $R_{\rm emit}$ is given in pc, and all component fluxes are in units of $10^{-14}\,\mathrm{TeV\,cm^{-2}\,s^{-1}}$. Only $R_{\rm emit}$ is varied; all other parameters are those
of Case 8 in Table \ref{tab:parameter_scan}. The $R_{\rm emit}=12.5\,\mathrm{pc}$ row reproduces that case. Component fluxes follow the same simultaneous spatial decomposition described in Subsect. \ref{app:C}. Values in parentheses give each component's percentage contribution to the total fitted flux.}}
\end{table*}

\begin{figure*}
    \centering
    \raisebox{0.8cm}{\includegraphics[width=0.49\textwidth]{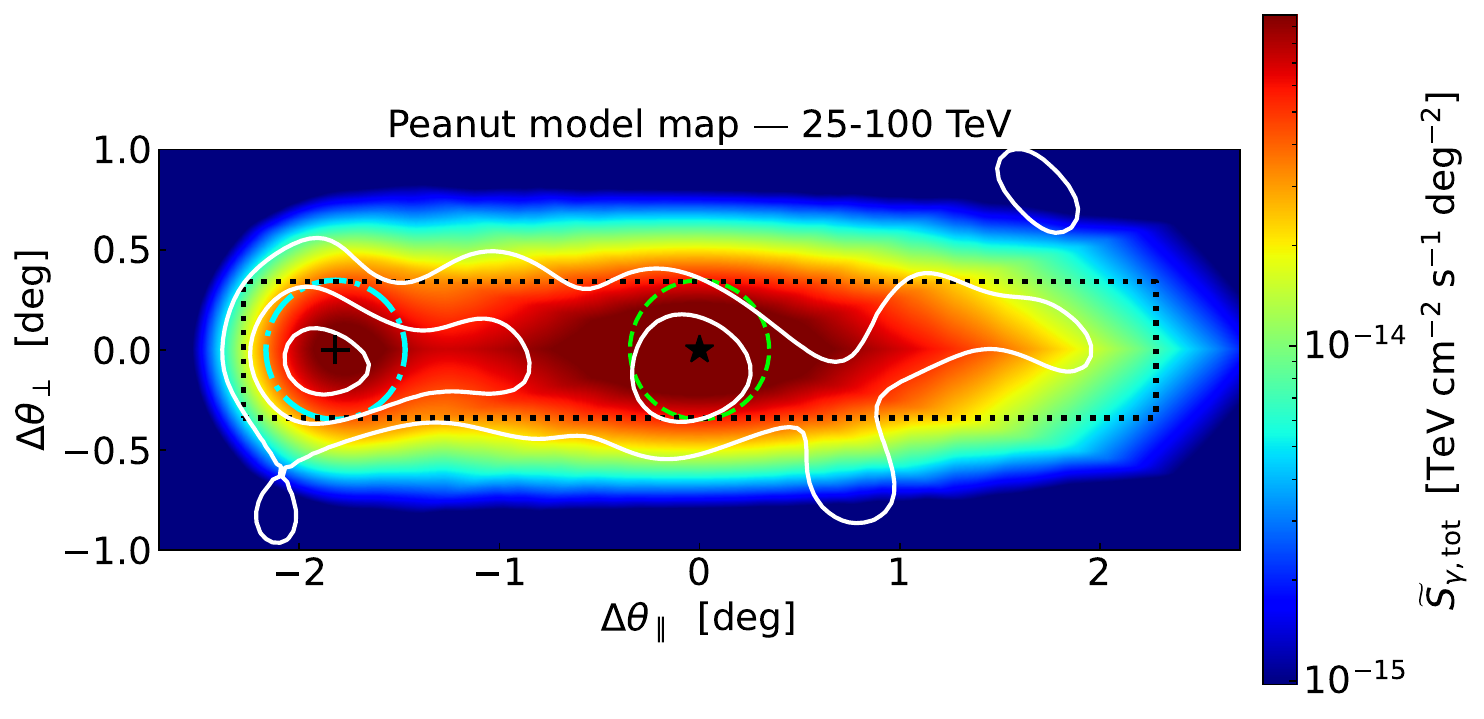}}
    \raisebox{0.8cm}{\includegraphics[width=0.49\textwidth]{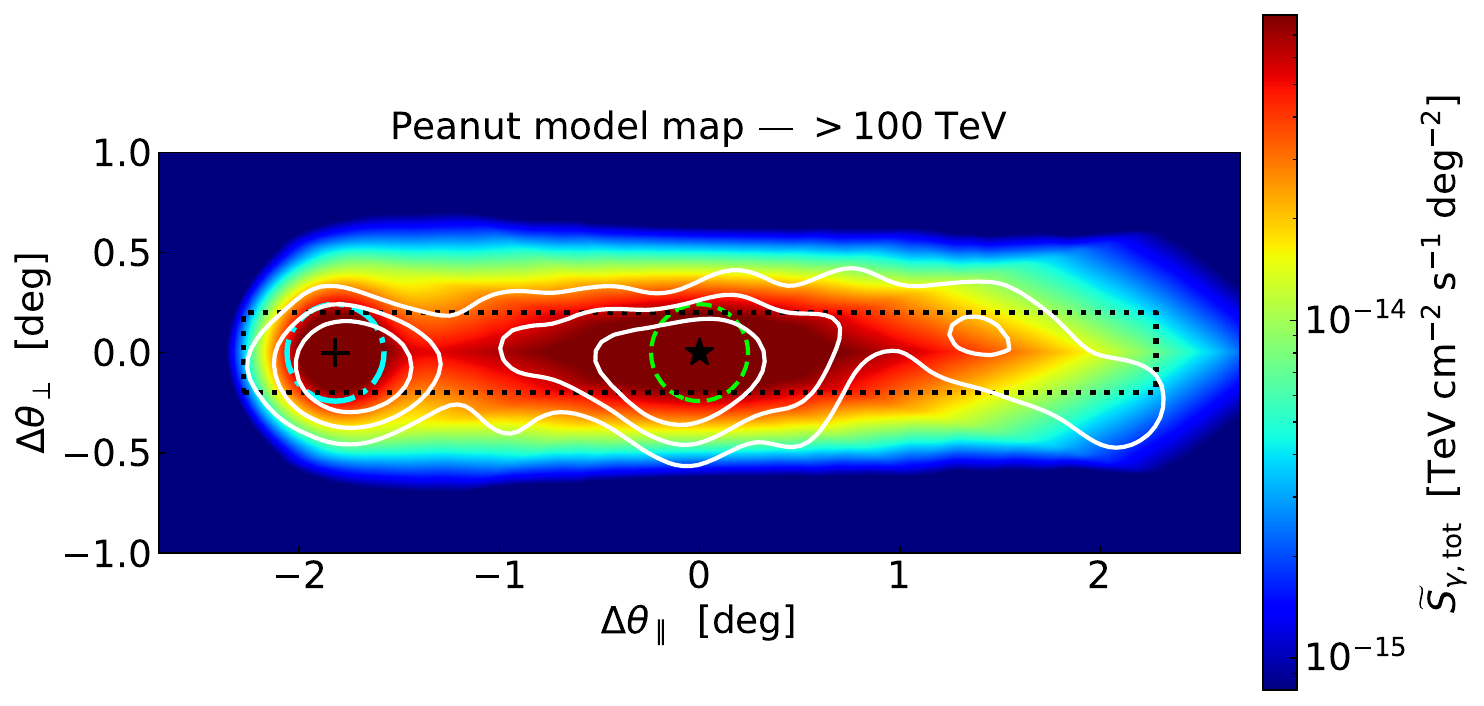}}
    \raisebox{0.8cm}{\includegraphics[width=0.49\textwidth]{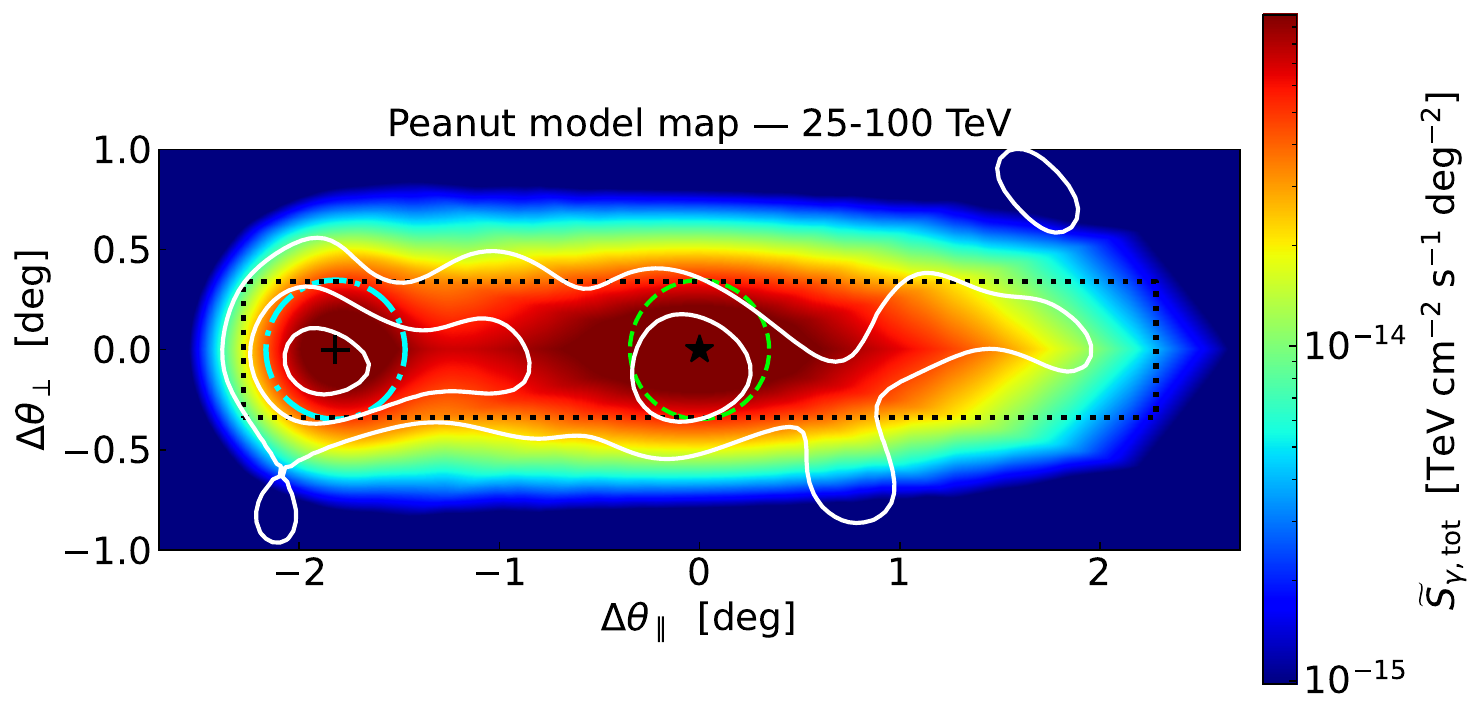}}
    \raisebox{0.8cm}{\includegraphics[width=0.49\textwidth]{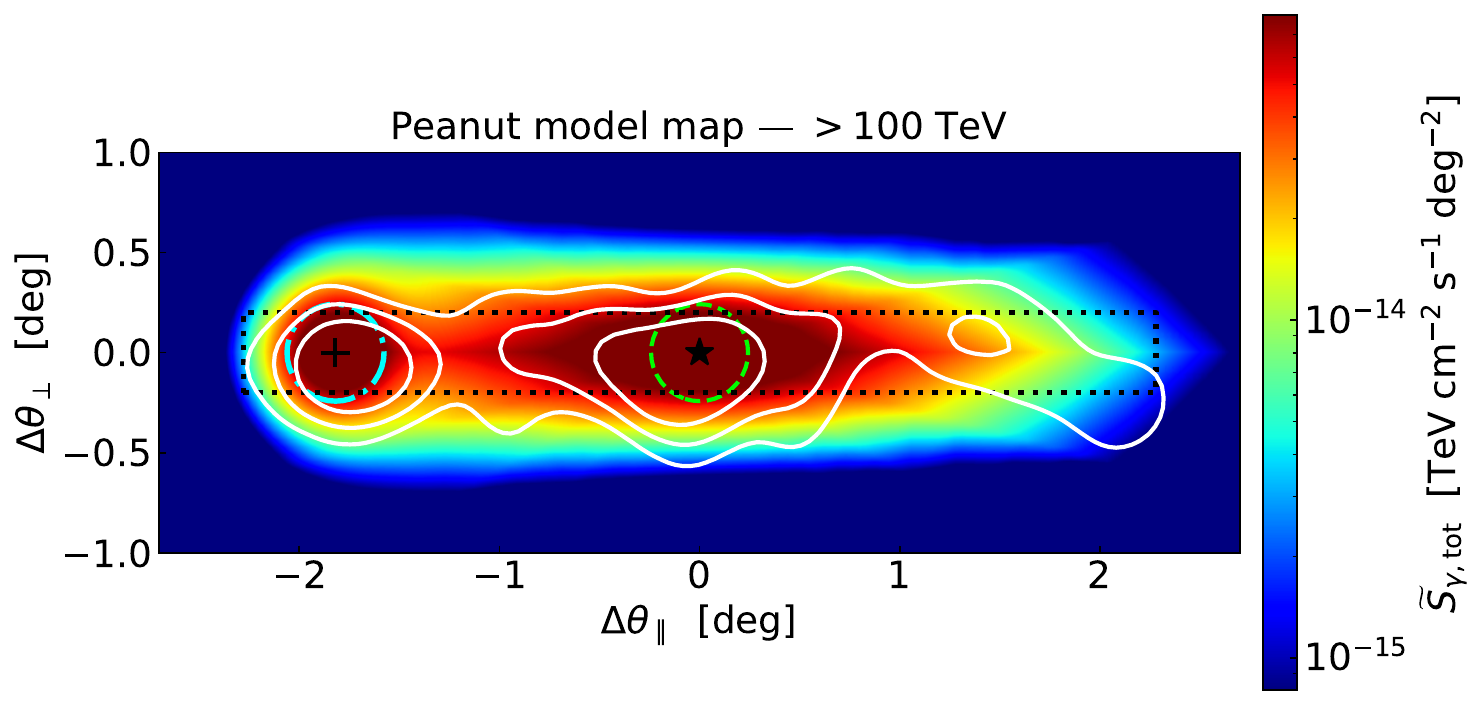}}
     \raisebox{0.8cm}{\includegraphics[width=0.49\textwidth]{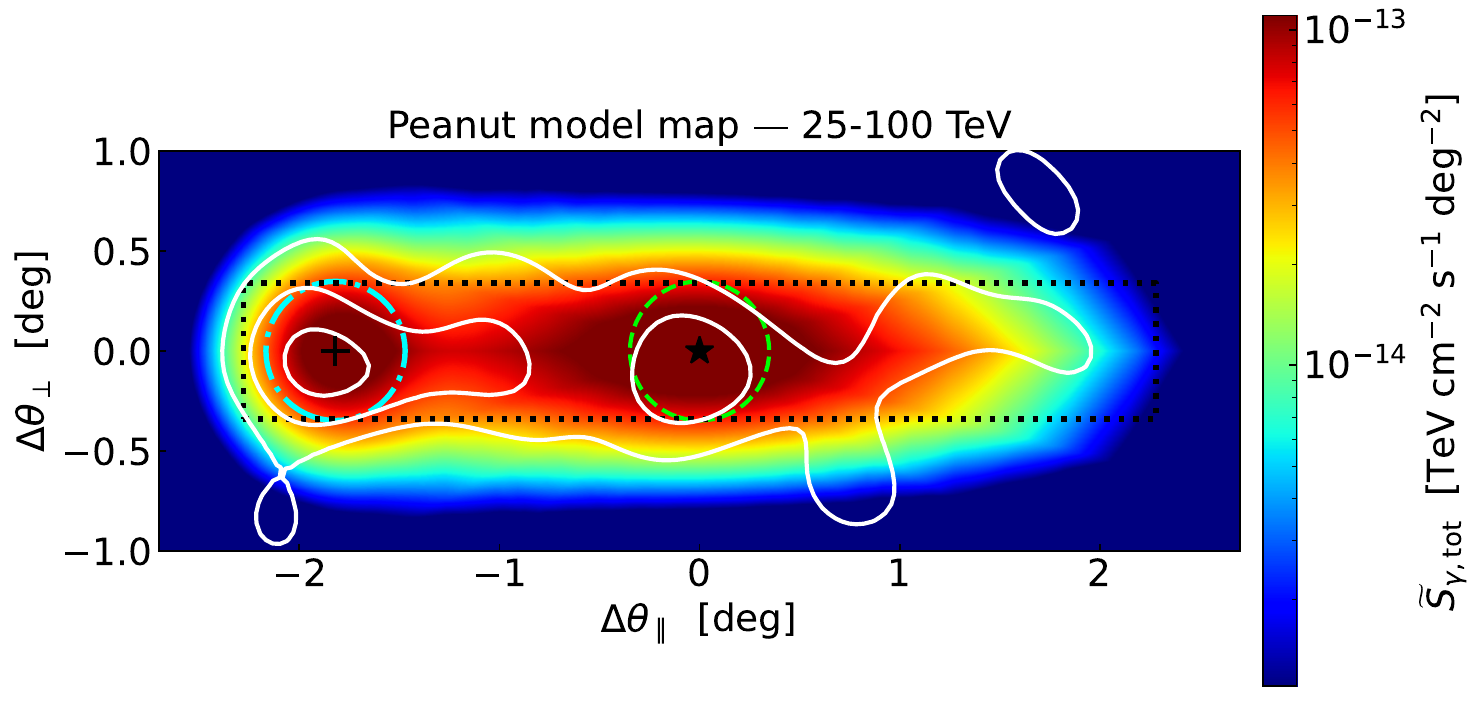}}
    \raisebox{0.8cm}{\includegraphics[width=0.49\textwidth]{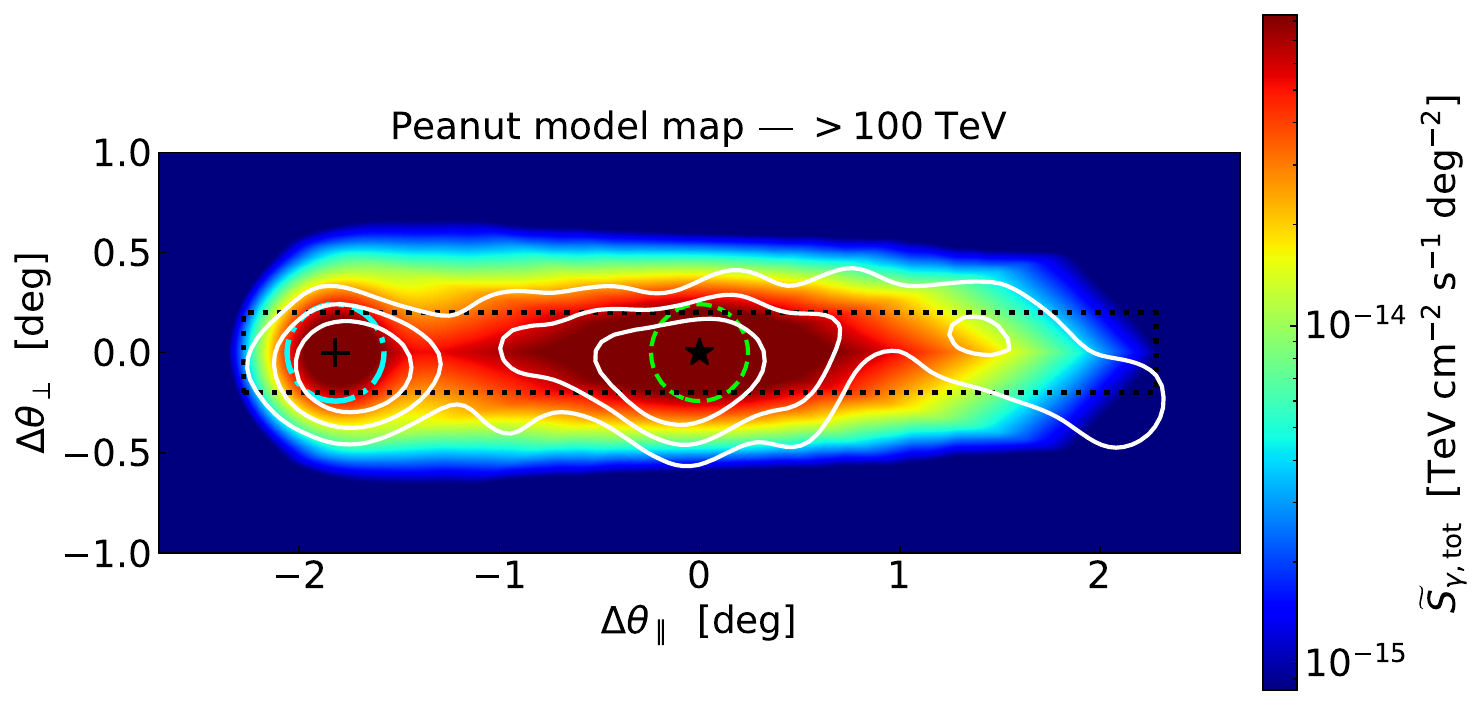}}
    \caption{Total model SB maps of the Peanut in the 25-100 TeV {(left column)} and $> 100 \ {\rm TeV}$ (right column) bands. {The rows show Cases 4, 8, and 12 of Table \ref{tab:parameter_scan}, with source ages of $10$, $20$, and $30\,\mathrm{kyr}$, from top to bottom.} The color scales show the {PSF-convolved} total model emission {after replacement of the original downstream contribution by the reoriented column.} White solid contours show the LHAASO significance levels at $3$, $5$, and $7\sigma$. The black dotted rectangle shows the intrinsic, band-dependent MD4 rectangular geometries adopted for the diffuse strip component \citep{lhaaso25}. Green dashed and cyan dash-dotted circles show the PSF-convolved central and far point source templates, respectively, {used in} the simultaneous MD4 decomposition. For each compact component, the circle is drawn at the $68\%$ containment radius of the adopted Gaussian PSF. The black star marks the position of the hidden central accelerator, and the black plus sign marks the {modeled} far hotspot position. {Coordinates are angular offsets along and across the modeled strip.}}
    \label{fig: SBmap}
\end{figure*}

For the injected particle spectrum, we chose a spectral shape given by Eq. \ref{eq:injection_appendix}, which is further normalized by Eq. \ref{eq:injection_normalization}, {over the electron energy interval $100\,\mathrm{GeV}$-$1.7\,\mathrm{PeV}$.} {We retain an injection index $p=1.5$, consistent with the hard spectra considered in pulsar halo models \citep{wu24,desarkar26}, and a cutoff-shape parameter $\beta=1$. The cutoff energy and conversion efficiency are adjusted across the explored configurations to reproduce the observed SED, as listed in Table \ref{tab:parameter_scan}.} {For the representative $t_{\rm age} = 20\,\mathrm{kyr}$ case, a cutoff energy of $E_c=420\,\mathrm{TeV}$ and conversion efficiency of $\eta=6.2\times10^{-4}$ are required.} For a discussion of whether this hidden putative pulsar-PWN system can accelerate particles to PeV energies, we refer to Appendix \ref{app:hidden_acc}. To calculate the IC emission, we consider the radiation field only from the CMB with the parameters $T_{\rm CMB} = 2.73 \ \rm K$ and $u_{\rm CMB} = 0.26 \ \rm eV \ cm^{-3}$. We retain only the CMB in the fiducial IC calculation, since infrared and optical fields contribute only a few percent in the relevant energy range because of strong KN suppression. {Figure \ref{fig: spectra} shows representative model spectra for the three source ages, together with the LHAASO measurements.} We find that the total model SED obtained within the source aperture, {which includes the baseline anisotropic halo after replacement of the original downstream contribution by the reoriented column}, can explain the observed spectral data points for the entire Peanut. {These examples remain compatible with the observed broad spectral shape after adjustment of the injection cutoff and efficiency.} Additionally, to obtain the PSF-convolved model SB maps, we convolved the model predictions with a symmetric two-dimensional Gaussian PSF. In this work, the width $\sigma_{\rm PSF}$ of the 2D Gaussian PSF is considered to be the same as LHAASO observations, i.e., $\sigma_{\rm PSF} = 0.23^{\circ}$ in 25-100 TeV and $\sigma_{\rm PSF} = 0.16^{\circ}$ in $>100$ TeV energy bands.

The component fluxes are obtained by simultaneously fitting the final SB maps with the LHAASO MD4 spatial templates, consisting of a uniform rectangular strip and two PSF-convolved point-source components, {as described in Subsect. \ref{app:C}.} Each spatial template is normalized to unit integral, so its fitted normalization directly corresponds to the band-integrated flux assigned to that component. {For the representative $t_{\rm age} = 20\,\mathrm{kyr}$ configuration, Case 8 in Table \ref{tab:parameter_scan},} the total band-integrated fluxes represented by the three fitted MD4 components are approximately $F_{\rm tot} = 2.452\times10^{-13} \, {\rm TeV\,cm^{-2}\,s^{-1}}$ in the 25-100 TeV band and $F_{\rm tot} = 1.512\times10^{-13} \, {\rm TeV\,cm^{-2}\,s^{-1}}$ in the $> 100$ TeV band. {Note that these flux values refer to the morphological decomposition.} The diffuse strip remains the dominant component, contributing $F_{\rm strip} = 1.841\times10^{-13} \, {\rm TeV\,cm^{-2}\,s^{-1}}$, or $75.10\%$, in 25-100 TeV and $F_{\rm strip} = 1.166\times10^{-13} \, {\rm TeV\,cm^{-2}\,s^{-1}}$, or $77.12\%$, above $100$ TeV. These fractions are close to the approximately \(70\%\) diffuse contribution inferred observationally. {The central excess contributes $F_{\rm c}=3.700\times10^{-14} \, {\rm TeV\,cm^{-2}\,s^{-1}}$ and $1.987\times10^{-14} \, {\rm TeV\,cm^{-2}\,s^{-1}}$ in the two bands, corresponding to $15.09\%$ and $13.14\%$ of the fitted flux, respectively. The corresponding far excess fluxes are $F_{\rm f}=2.406\times10^{-14} \, {\rm TeV\,cm^{-2}\,s^{-1}}$ and $1.473\times10^{-14} \, {\rm TeV\,cm^{-2}\,s^{-1}}$, or $9.81\%$ and $9.74\%$, respectively. The resulting far-to-central flux ratios are $F_{\rm f}/F_{\rm c}\approx 0.65$ and $0.74$ in the two energy bands.} {The compact components have fluxes of the same order, but the far component is comparatively weaker in both bands, while the strip contribution remains above the approximately $70\%$ inferred observationally.}
Fig. \ref{fig: SBmap} shows the total model {PSF-convolved} SB maps, i.e., $\widetilde S_{\gamma,{\rm tot}} = \widetilde S_{\gamma,{\rm base}} - \widetilde S_{\gamma,{\rm rem}} + \widetilde S_{\gamma,{\rm column}},$ in the \(25\)-\(100\) TeV and $>100$ TeV bands, {replacing the original downstream contribution with that of the remapped electron column.} The total emission maps retain the dominant elongated narrow strip and show a bright enhancement at {the central excess position, together with a comparatively weaker enhancement at the far excess.}

{The wider exploration in Table \ref{tab:parameter_scan} shows that the relative strength of the far component depends on the adopted transport and magnetic field parameters. For $D_{\parallel,0}=3.2\times10^{29}\, \mathrm{cm^2\,s^{-1}}$, its fitted fraction ranges from
$0$ to approximately $5.1\%$ across the explored ages, field strengths, and energy bands, while the diffuse strip contributes approximately $81$-$84\%$. For $D_{\parallel,0}=3.2\times10^{30}\,
\mathrm{cm^2\,s^{-1}}$, the far fraction increases to approximately $6.0$-$12.3\%$, with a diffuse-strip contribution of approximately $73$-$80\%$. Thus, within the explored configurations, faster parallel transport produces a stronger far component and brings the component brightness division closer to that inferred observationally. Nevertheless, even in configurations where the far and central components have comparable fitted fluxes at 25-100 TeV, the far component remains weaker above 100 TeV, and complete agreement with the observed component partition is not established. Note that the zero far component fraction obtained in the higher-energy band of Case 11 has the interpretation described in Subsect. \ref{app:C}, i.e., the simultaneous decomposition assigns no additional positive flux to the far point source template. It does not imply that the downstream electron population or the emission from its reoriented column vanishes.

We retain Case 8, with $t_{\rm age}=20\,\mathrm{kyr}$, $B=0.1\,\mu\mathrm{G}$, and $D_{\parallel,0}=3.2\times10^{30}\, \mathrm{cm^2\,s^{-1}}$, as a representative example. Its far component contributes approximately $10\%$ in each band, while the central component contributes approximately $13$-$15\%$. The exploration does not uniquely favor a weak ambient magnetic field, as the higher magnetic field configurations also produce appreciable far emission, with its relative strength depending on the energy band and source age. The representative SEDs remain compatible with the observed spectrum after adjustment of the injection normalization and cutoff energy. The spectra and maps do not, however, establish a unique source age, so the $20\,\mathrm{kyr}$ example is used as a reference configuration rather than a measured age.

Table \ref{tab:emit_radius_scan} examines the dependence on the emitting radius for Case 8 while keeping the selected downstream population fixed. The fitted far fraction decreases as $R_{\rm emit}$ increases. Increasing $R_{\rm emit}$ from $6.25$ to $50\,\mathrm{pc}$ reduces the fitted far fraction from $10.00\%$ to $6.92\%$ in the $25$-$100\,\mathrm{TeV}$ band and from $10.15\%$ to $4.58\%$ in $>100$ TeV band. Even for the largest emitting radius, a nonzero far component remains in both bands, although a smaller fraction of the emission is assigned to the far point source template.
The difference between $R_{\rm emit}=6.25$ and $12.5\,\mathrm{pc}$ is much smaller. Further concentration therefore provides little additional enhancement in the adopted point source decomposition once the emission is sufficiently compact relative to the PSF. Throughout this radius exploration, the electron population, radiation field, and column distance remain fixed. Changing $R_{\rm emit}$ therefore changes the spatial
distribution of the emission without changing its total IC luminosity. The decrease in the fitted far fraction reflects the reduced apparent compactness of the broader column.

{For the representative Case~8, the parallel scattering mean free path, $\lambda_\parallel=3D_\parallel/c$, is approximately $104\,\mathrm{pc}$ at $100\,\mathrm{TeV}$ and
$164\,\mathrm{pc}$ at $400\,\mathrm{TeV}$. In the adopted ambient field of $B=0.1\,\mu\mathrm{G}$,
a $1\,\mathrm{PeV}$ electron has a Larmor radius of approximately $11\,\mathrm{pc}$, comparable to the emitting column radius $R_{\rm emit}=12.5\,\mathrm{pc}$. The diffusion approximation may therefore have limited accuracy under these conditions, particularly near the assumed magnetic field reorientation. These limitations may become more important if $D_\parallel$ is increased substantially beyond the explored range, as particles would scatter less frequently and their propagation could become more sensitive to the local magnetic field geometry.} Such conditions may produce filamentary structures or localized projected enhancements \citep{bao25a,bao25b}, although these features are not determined by the diffusion coefficient alone. Our results therefore identify the limitations of the explored configurations, without establishing that reproducing the observed far excess necessarily requires transport outside the diffusive regime.

Taken together, these results show that the same propagated electron population can generate a dominant strip, a central concentration, and a far enhancement under the adopted field geometry. The relative brightness of the far compact component, particularly at higher energy band, nevertheless remains a limitation of the explored particle-conserved remapping prescription. }

\section{Concluding remarks}
\label{sec:conclusion}

We investigated whether the LHAASO Peanut can be explained by particle injection from a single hidden central accelerator followed by anisotropic transport and a local magnetic field reorientation. {Although PSR J0218+4232 is the most obvious cataloged object in the region, its proper motion cannot account for the observed extent, and a possible BSPWN faces demanding acceleration, confinement, and escape requirements according to the simple estimates discussed in this paper. The short cooling time of the emitting electrons requires recent or ongoing injection, rather than a fossil population accumulated over the MSP lifetime.} We therefore considered a hidden pulsar-PWN system near the central excess. {The parameter exploration covers source ages of $10$, $20$, and $30\,\mathrm{kyr}$, together with different magnetic field strengths and parallel diffusion coefficients. The viewing geometry is constrained by the observed dimensions using an effective propagation time limited by both source age and radiative cooling. Representative configurations describe the broad observed SED and produce central and far components in the adopted spatial decomposition. Increasing the parallel diffusion coefficient strengthens the far component, bringing its fitted flux close to that of the central component in the 25-100 TeV band in some configurations, although it remains lower above 100 TeV. In the representative $20\,\mathrm{kyr}$ case, the strip contributes approximately $75$-$77\%$ of the fitted flux, the central component contributes approximately $13$-$15\%$, and the far component contributes approximately $10\%$.

The baseline halo supplies the downstream electrons, whose particle-conserving redistribution into a favorable LOS-aligned column enhances the far emission while replacing their original contribution. Varying the emitting radius changes the compactness and fitted far fraction, but does not increase the total IC luminosity of the selected population. Further concentration below the reference emitting radius produces only a small additional change in the fitted far component. The $t_{\rm age} = 20\,\mathrm{kyr}$ configuration provides a representative illustration, as the present exploration does not uniquely determine the source age.}
{The local magnetic field reorientation and particle-conserved remapping remain a phenomenological plausibility test rather than a unique magnetic field reconstruction or a self-consistent three-dimensional field deformation \citep{bao25a,bao25b}.} {The calculation therefore demonstrates that a single central accelerator can produce a far enhancement within the adopted framework, although the explored configurations do not fully reproduce its observed brightness relative to the other components, especially at higher energies.} Deep radio and X-ray observations of the central region, improved energy-resolved measurements of the compact components, and constraints on the local field geometry will be {important in testing this interpretation}. In particular, a relatively brighter X-ray strip \citep{bao25a}, or a compact X-ray PWN associated with a radio-quiet central pulsar would support the proposed interpretation.

\begin{acknowledgements}
      A.D.S. thanks the anonymous reviewer for helpful suggestions and constructive criticism, which vastly improved the quality of this work. 
      A.D.S. is supported by the grant Juan de la Cierva JDC2023-052168-I, funded by MCIU/AEI/10.13039/501100011033 and by the ESF+.
      This work has been supported by the grant PID2024-155316NB-I00 funded by MICIU /AEI /10.13039/501100011033 / FEDER, UE, and CSIC PIE 202350E189. This work is also partly supported by the Spanish program Unidad de Excelencia María de Maeztu CEX2020-001058-M, financed by MCIN/AEI/10.13039/501100011033, and by the MaX-CSIC Excellence Award MaX4-SOMMA-ICE, and also supported by MCIN with funding from European Union NextGeneration EU (PRTR-C17.I1).
\end{acknowledgements}

%
   \bibliographystyle{aa} 
   \bibliography{aa.bib} 

\appendix

\section{Distance dependence of the model parameters}
\label{app:distance}

{The distance to the putative hidden pulsar is presently unconstrained. We adopt $d_0=3.15\,\mathrm{kpc}$ to facilitate comparison with the LHAASO analysis, without implying a physical association with PSR J0218+4232. Here we describe the distance dependence using the representative $t_{\rm age}=20\,\mathrm{kyr}$ configuration. The same relations can be applied separately to the other configurations in Table \ref{tab:parameter_scan}.

We define
\begin{equation}
f_d=\frac{d}{d_0}=\frac{d}{3.15\,\mathrm{kpc}}.
\label{eq:distance_factor}
\end{equation}
For fixed observed angular dimensions, the projected physical
scales vary linearly with distance:
\begin{equation}
\begin{aligned}
R_{\perp,{\rm obs}}(d)
&=R_{\perp,{\rm obs}}(d_0)f_d,\\
R_{\parallel,{\rm obs}}(d)
&=R_{\parallel,{\rm obs}}(d_0)f_d.
\end{aligned}
\label{eq:physical_distance_scaling}
\end{equation}
At the reference distance, these scales are approximately 12.5 and 125 pc, respectively. Here, $R_{\perp, \rm obs}$ and $R_{\parallel, \rm obs}$ denote the projected transverse half-width and longitudinal half-scale of the halo, respectively.

We retain the source age, ambient magnetic field, radiation field, and parallel diffusion coefficient of the reference configuration, i.e.,
$t_{\rm age}=20\,\mathrm{kyr}$,
$B=0.1\,\mu\mathrm{G}$, and
$D_{\parallel,0}=3.2\times10^{30}\,
\mathrm{cm^2\,s^{-1}}$.
The effective propagation time evaluated at
$E_*=400\,\mathrm{TeV}$ therefore remains
$t_{\rm eff}=20\,\mathrm{kyr}$.
At $d_0$, the remaining parameters are
$D_{\perp,0}(d_0)=3.652\times10^{26}\,
\mathrm{cm^2\,s^{-1}}$,
$\Phi_0=6.240^\circ$, and $M_A(d_0)=0.1034$.

Preserving the transverse angular extent in Eq. \ref{eq:morphology_scaling} requires
\begin{equation}
D_{\perp,0}(d)
=
D_{\perp,0}(d_0)f_d^2.
\label{eq:dperp_distance}
\end{equation}
Retaining both diffusion components in the projected longitudinal extent gives
\begin{equation}
\sin\Phi(d)
=
f_d\sin\Phi_0
\left[
\frac{D_{\parallel,0}-D_{\perp,0}(d_0)}
{D_{\parallel,0}-D_{\perp,0}(d_0)f_d^2}
\right]^{1/2}.
\label{eq:phi_distance}
\end{equation}
Since the coefficients have the same energy dependence, their ratio can be evaluated at the common reference energy. For strong anisotropy, Eq. \ref{eq:phi_distance} reduces to $\sin\Phi(d)\approx f_d\sin\Phi_0$.
The Alfv\'enic Mach number scales as
\begin{equation}
M_A(d)=M_A(d_0)f_d^{1/2},
\label{eq:ma_distance}
\end{equation}
following from
$M_A=(D_{\perp,0}/D_{\parallel,0})^{1/4}$.

For a representative distance range $d=2$-$4\,\mathrm{kpc}$, these relations give
\begin{equation}
D_{\perp,0}
\approx
(1.47\text{-}5.89)\times10^{26}\,
\mathrm{cm^2\,s^{-1}},
\end{equation}
and
\begin{equation}
\Phi\approx3.96^\circ\text{-}7.93^\circ,
\qquad
M_A\approx0.082\text{-}0.116.
\end{equation}
The parallel diffusion coefficient remains fixed by construction, and transport remains strongly anisotropic throughout this interval.

For the fixed angular separation $\alpha=1.82^\circ$, the downstream selection boundary and the distance to the reorientation point follow from the observer geometry,
\begin{equation}
\begin{aligned}
L_{\rm sample}(d)
&=
\frac{d\sin\alpha}{\sin[\Phi(d)+\alpha]},\\
d_{\rm bend}(d)
&=
\frac{d\sin\Phi(d)}{\sin[\Phi(d)+\alpha]}.
\end{aligned}
\end{equation}
At $d_0=3.15\,\mathrm{kpc}$, these distances are approximately $714\,\mathrm{pc}$ and $2.44\,\mathrm{kpc}$, respectively. Over $d=2$-$4\,\mathrm{kpc}$, with $\Phi(d)$ determined by Eq. \ref{eq:phi_distance}, $L_{\rm sample}$ varies from approximately $631$ to $750\,\mathrm{pc}$, while $d_{\rm bend}$ varies from approximately $1.37$ to $3.26\,\mathrm{kpc}$.

The projected electron column is normalized by the downstream population according to Subsect. \ref{app:reoriented_column}. If its angular radius is retained when varying distance, its physical emitting radius scales as
\begin{equation}
R_{\rm emit}(d)
=
12.5\,
\frac{d_{\rm bend}(d)}{d_{\rm bend}(d_0)}
\,\mathrm{pc}.
\end{equation}
This gives approximately $7.0$-$16.7\,\mathrm{pc}$ over $d=2$-$4\,\mathrm{kpc}$. Preserving the angular source-centered aperture instead requires its physical radius to scale as $155f_d\,\mathrm{pc}$.
The downstream population and its emission must be evaluated for the corresponding transport and observer geometry, whereas the component fractions are not fixed by these geometrical relations alone.

The source energetics are more directly sensitive to distance. Assigning the source-integrated emission to the accelerator distance gives the usual isotropic-equivalent luminosity estimate,
\begin{equation}
L_\gamma=4\pi d^2F_\gamma\propto d^2.
\label{eq:luminosity_distance}
\end{equation}
To first order, if the intrinsic pulsar properties, injection spectral shape, radiative environment, and fraction of emission enclosed by the angular aperture remain unchanged, the required conversion efficiency scales as
\begin{equation}
\eta(d)\approx\eta(d_0)f_d^2.
\label{eq:eta_distance}
\end{equation}
For the reference value $\eta(d_0)=6.2\times10^{-4}$, this gives approximately $\eta=2.5\times10^{-4}$ at $2\,\mathrm{kpc}$ and $\eta=1.0\times10^{-3}$ at $4\,\mathrm{kpc}$. A smaller distance therefore relaxes the particle energy requirement, whereas a larger distance makes it comparatively more demanding. The maximum particle energy allowed by the pulsar spin-down power, discussed in Appendix \ref{app:hidden_acc}, is unchanged by distance if the intrinsic pulsar parameters are kept fixed. This is a leading estimate of the required injected power. The reoriented contribution also depends on $d_{\rm bend}$, so the exact distance dependence of the total emission and component fractions requires recalculation.

Finally, neglecting the Solar height above the Galactic plane, the Galactic height of the source scales linearly with distance,
\begin{equation}
|z_{\rm Gal}|
=
d|\sin b|
\approx0.95f_d\,\mathrm{kpc}.
\label{eq:zgal_distance}
\end{equation}
The interval $d=2$-$4\,\mathrm{kpc}$ therefore corresponds to $|z_{\rm Gal}|\approx0.6$-$1.2\,\mathrm{kpc}$. Distance uncertainty changes the inferred physical size, transverse transport, viewing angle, and energy requirement. The relations above describe these dependencies about the reference configuration, whereas the detailed spectrum and component fluxes require the corresponding transport and emission calculation at each distance.}

\section{PeV electron acceleration by the hidden pulsar-PWN system}
\label{app:hidden_acc}

{The considerations presented in Sect. \ref{app: MSP}, together with the detection of the highest-energy photon near the central excess, motivate an unidentified young or middle-aged compact pulsar-PWN system near that position. Here we examine whether the adopted present-day spin-down power of the putative pulsar can support electron acceleration into the PeV regime. We use the representative $t_{\rm age} = 20\,\mathrm{kyr}$ configuration for the detailed example and then consider the other source ages in Table \ref{tab:parameter_scan}.

The reference configuration has an exponential injection cutoff of $E_{\rm c}=420\,\mathrm{TeV}$, while the explored configurations span $E_{\rm c}=300$-$850\,\mathrm{TeV}$. An exponential cutoff does not represent a sharp maximum energy, and electrons remain present above $E_{\rm c}$. In all cases, the electron spectrum is followed up to $E_{\max}=1.7\,\mathrm{PeV}$. Because the observed UHE gamma-ray emission requires
electrons with energies of order PeV, we examine the illustrative requirement $E_{\rm acc,max}\gtrsim1.7\,\mathrm{PeV}$. This tests the ability to supply the modeled PeV energy tail, and is not a uniquely measured cutoff of the accelerator.

Following \cite{lhaaso26}, the maximum particle energy supported by the electromagnetic potential of a pulsar wind is
\begin{equation}
E_{\rm H}
=
e\left(\frac{2\beta_{\rm w}\epsilon_B\dot E}{c}\right)^{1/2}
\approx
7.8
\left(
\frac{\beta_{\rm w}\epsilon_B\dot E}
{10^{37}\,\mathrm{erg\,s^{-1}}}
\right)^{1/2}
\,\mathrm{PeV}.
\label{eq:wind_potential_hidden}
\end{equation}
Here $\epsilon_B$ is the fraction of the pulsar wind power carried by the magnetic field in the acceleration region, and $\beta_{\rm w}$ is the wind speed in units of $c$. We adopt $\beta_{\rm w}\approx1$ for an ultrarelativistic pulsar wind. The practically accessible energy is reduced relative to this limiting potential by the acceleration efficiency,
\begin{equation}
E_{\rm acc,max}^{\rm pot}
=
\eta_{\rm acc}E_{\rm H},
\label{eq:emax_potential_hidden}
\end{equation}
where $\eta_{\rm acc}=\mathcal{E}_{\rm eff}/B_{\rm acc}$, $\mathcal{E}_{\rm eff}$ is the effective accelerating electric field averaged along the particle trajectory, and $B_{\rm acc}$ is the characteristic magnetic field in the acceleration region. The acceleration efficiency $\eta_{\rm acc}$ is distinct from the pair conversion efficiency $\eta$ in Eq. \ref{eq:injection_normalization}. Likewise, $B_{\rm acc}$ describes the compact accelerator and need not equal the ambient magnetic field governing transport through the extended halo.

For the adopted present-day spin-down luminosity of the $t_{\rm age} = 20\,\mathrm{kyr}$ configuration, $\dot E=3.0\times10^{36}\,\mathrm{erg\,s^{-1}}$, the potential limit becomes
\begin{equation}
E_{\rm acc,max}^{\rm pot}
\approx
4.27\,\eta_{\rm acc}\epsilon_B^{1/2}
\,\mathrm{PeV}.
\end{equation}
Requiring electrons to reach at least $1.7$ PeV therefore gives
\begin{equation}
\eta_{\rm acc}
\gtrsim
0.40\,\epsilon_B^{-1/2}.
\label{eq:eta_potential_hidden}
\end{equation}
For $\eta_{\rm acc} < 1$, the potential constraint alone therefore requires $\epsilon_B\gtrsim0.16$. Representative magnetizations $\epsilon_B=0.3$-$0.8$ correspond to minimum acceleration efficiencies decreasing from approximately $0.73$ to $0.45$. The adopted spin-down power can thus support PeV acceleration, but requires an efficient and substantially magnetized acceleration region. These parameters therefore describe an extreme PeV accelerator rather than a typical weakly magnetized termination shock environment. For comparison, \cite{lhaaso26} found a minimum acceleration efficiency of order $0.3$ for PSR J1849-0001, whose spin-down luminosity is substantially larger than that adopted here.

The available electric potential is not the only limitation. For electrons, synchrotron cooling provides an independent upper bound. Following \cite{lhaaso26}, the magnetic field in the acceleration region can be related to the pulsar wind power through
\begin{equation}
\frac{B_{\rm acc}^2}{8\pi}
=
\frac{\epsilon_B\dot E}
{4\pi f_\omega R_{\rm acc}^2c},
\label{eq:acceleration_field_hidden}
\end{equation}
where $R_{\rm acc}$ is the characteristic acceleration region radius and $f_\omega$ is the equivalent filling factor of the pulsar wind. We adopt $f_{\omega}=1$, following the fiducial treatment of \cite{lhaaso26}. Equating the acceleration rate,
$\dot E_{\rm acc}
=
e\mathcal{E}_{\rm eff}c
=
\eta_{\rm acc}eB_{\rm acc}c,$
with the synchrotron loss rate then gives
\begin{equation}
E_{\rm acc,max}^{\rm syn}
\approx
6.3\,\eta_{\rm acc}^{1/2}
\left(\frac{R_{\rm acc}}{0.1\,\mathrm{pc}}\right)^{1/2}
\times
\left(
\frac{\epsilon_B\dot E}
{10^{37}\,\mathrm{erg\,s^{-1}}}
\right)^{-1/4}
\,\mathrm{PeV}.
\label{eq:synch_acceleration_hidden}
\end{equation}
Within this description, the physically attainable maximum electron energy is therefore bounded by
\begin{equation}
E_{\rm acc,max}
=
\min\left(
E_{\rm acc,max}^{\rm pot},
E_{\rm acc,max}^{\rm syn}
\right).
\end{equation}

For the illustrative choice $\epsilon_B=0.5$, $\eta_{\rm acc}=0.57$, and $R_{\rm acc}=0.01\,\mathrm{pc}$, the potential and synchrotron limits are approximately $E_{\rm acc,max}^{\rm pot}\approx1.7\ {\rm PeV} \ \text{and} \ E_{\rm acc,max}^{\rm syn}\approx2.4 \ {\rm PeV},$ respectively, so that the maximum energy is set by the available potential rather than by synchrotron cooling. The corresponding field is $B_{\rm acc}\approx3.2\times10^2\,\mu\mathrm{G}$. Thus, the weak ambient halo field used in the transport model is compatible with a much stronger field in the compact acceleration region. Here, note that although both the MSP/BSPWN interpretation discussed in Sect. \ref{app: MSP} and the hidden central pulsar-PWN interpretation involve compact acceleration regions, the larger spin-down power adopted for the hidden pulsar provides a higher available potential. The synchrotron constraint is imposed in both cases. More generally, for a representative set of paired magnetizations and efficiencies,
\begin{equation}
\begin{aligned}
\epsilon_B&\approx0.3-0.8,\\
\eta_{\rm acc}&\approx0.45-0.73,\\
R_{\rm acc}&=10^{-2}\,\mathrm{pc},
\end{aligned}
\label{eq:allowed_acc_range}
\end{equation}
the potential constraint permits $1.7$ PeV electrons and the synchrotron loss limit remains above the required energy. At these efficiencies, the synchrotron limit, $E_{\rm acc,max}^{\rm syn}$, decreases from
approximately $3.1\,\mathrm{PeV}$ at $\epsilon_B=0.3$ to $1.9\,\mathrm{PeV}$ at $\epsilon_B=0.8$, remaining above the required $1.7$ PeV energy throughout this representative interval. At these paired efficiencies and magnetizations, the minimum radii imposed by the synchrotron constraint range from approximately $0.003$ to $0.008\,\mathrm{pc}$. The adopted $0.01\,\mathrm{pc}$ radius is therefore a sufficient illustrative scale rather than a uniquely inferred size.

A further constraint follows from the finite pulsar spin-down budget. When the effective accelerating electric field becomes a substantial fraction of the magnetic field, its associated energy flux is no longer
negligible. The combined electric and magnetic luminosities must therefore satisfy
\begin{equation}
\left(1+\eta_{\rm acc}^2\right)\epsilon_B<1.
\label{eq:acceleration_budget}
\end{equation}
For $\epsilon_B=0.5$ and $\eta_{\rm acc}=0.57$, the left-hand side is approximately $0.66$. At the upper end of the representative interval, $\epsilon_B=0.8$ and $\eta_{\rm acc}\approx0.45$ give approximately $\left( 1+\eta_{\rm acc}^{2} \right)\epsilon_B \approx 0.96$, so this configuration lies close to the available electromagnetic spin-down power limit. Combining Eq. \ref{eq:eta_potential_hidden} with
Eq. \ref{eq:acceleration_budget} gives approximately $\epsilon_B\lesssim0.84$ for the adopted $t_{\rm age} = 20\,\mathrm{kyr}$ spin-down luminosity and a target energy of $1.7$ PeV. The interval in Eq. \ref{eq:allowed_acc_range} should therefore be regarded as a representative viable region rather than as a measured range, i.e., lower magnetization requires $\eta_{\rm acc}$ to approach or exceed unity, whereas the largest $\epsilon_B$ values approach the spin-down power budget.

The other explored ages can be assessed with the same relations. Their adopted present-day spin-down luminosities are approximately $6.3\times10^{36}$ and $1.8\times10^{36}\,\mathrm{erg\,s^{-1}}$
at $10$ and $30\,\mathrm{kyr}$, respectively. For fixed $\epsilon_B=0.5$ and $R_{\rm acc}=0.01\,\mathrm{pc}$, efficiencies $\eta_{\rm acc}\approx0.41$, $0.57$, and $0.74$ satisfy both the potential and synchrotron constraints at $10$, $20$, and $30\,\mathrm{kyr}$, respectively. All three choices also satisfy Eq. \ref{eq:acceleration_budget}. Consequently, none of the three adopted ages is excluded
by these acceleration bounds, although the lower spin-down power of the oldest configuration requires a higher acceleration efficiency. At a fixed age, changing the ambient magnetic field or diffusion coefficient does not directly change these compact accelerator bounds.

The resulting parameter space is physically allowed but extreme. This is qualitatively similar to the situation discussed by \cite{lhaaso26}, who showed that PeV acceleration at a weakly magnetized termination shock can require $\eta_{\rm acc}\gg1$, whereas acceleration in a more strongly magnetized region further upstream in the pulsar wind can substantially relax the required efficiency. Such regions may arise through dissipation of a highly magnetized wind or within compact compressed structures, rather than at the conventional termination shock itself. Thus, the estimates above show that a compact and sufficiently magnetized PWN acceleration region can supply electrons up to the energies required by the model without identifying a unique acceleration mechanism, although this necessarily implies an unusually efficient PeV accelerator.

The absence of a cataloged central pulsar or bright PWN does not exclude this interpretation. An evolved PWN may have been compressed and disrupted by the SNR reverse shock, promoting particle escape and leaving a faint, low-surface-brightness outer nebula \citep{bandiera23II,bandiera23III,matin24,martin25}. At the same time, the PeV acceleration considered here occurs in a much more compact region close to the pulsar, which would be unresolved at the adopted distance and need not correspond to a bright extended PWN. The detectability of the nebula may further decrease during post-compression evolution \citep{desarkar26a}, while unfavorable pulsar beaming can prevent a radio detection. At the adopted distance, the system lies at $|z_{\rm Gal}|\approx0.95$ kpc, making such a pulsar unusual and favoring either an off-plane progenitor or a smaller true distance. Deep radio and X-ray observations of the central excess provide a direct test of this scenario.}

\end{document}